\documentclass[journal]{IEEEtai}

\usepackage[colorlinks,urlcolor=blue,linkcolor=blue,citecolor=blue]{hyperref}

\usepackage{amssymb}
\usepackage{pifont}
\newcommand{\cmark}{\ding{51}}%
\newcommand{\xmark}{\ding{55}}%

\usepackage{makecell}

\usepackage[pdftex]{graphicx}
\usepackage{color,array}

\usepackage{graphicx}
\usepackage{cite}
\usepackage{amsmath,amssymb,amsfonts}
\usepackage{textcomp}
\usepackage{subfigure}
\usepackage[utf8]{inputenc}

\usepackage[ruled,vlined, norelsize]{algorithm2e}
\usepackage{booktabs}
\usepackage{pifont}
\usepackage[english]{babel}
\usepackage{multicol}
\usepackage{placeins}
\usepackage{amsmath}
\def\BibTeX{{\rm B\kern-.05em{\sc i\kern-.025em b}\kern-.08em
    T\kern-.1667em\lower.7ex\hbox{E}\kern-.125emX}}
  
\usepackage{tabu}
\usepackage{cleveref}

\begin{document}


\title{Evasion Generative Adversarial Network for Low Data Regimes}

\author{Rizwan Hamid Randhawa,
        Nauman Aslam,
        Mohammad Alauthman,
        Husnain Rafiq
        
\thanks{Manuscript received October XX, 2021; revised MONTH X, 2021 and MONTH XX, 2022; accepted MONTH X, 2022. Date of publication MONTH X, 2022; date of current version MONTH XX, 2022. This work was supported under the Research and Development Fund (RDF) of Northumbria University, Newcastle upon Tyne, UK.)
 }

\thanks{Rizwan Hamid Randhawa, Nauman Aslam and Husnain Rafiq are with the Department of Computer and Information Sciences, Northumbria University, Newcastle upon Tyne, UK (e-mail: {rizwan.randhawa, nauman.aslam, husnain.rafiq}@northumbria.ac.uk)}

\thanks{Mohammad Alauthman is with the Department of Information Security, University of Petra, Amman 11196, Jordan (e-mail: mohammad.alauthman@uop.edu.jo)}
}


\markboth{Accepted for publication in one of the future volumes of IEEE Transactions on Artificial Intelligence}
{Randhawa \MakeLowercase{\textit{et al.}}: Evasion Generative Adversarial Network for Low Data Regimes}

\maketitle

\begin{abstract}
A myriad of recent literary works has leveraged generative adversarial networks (GANs) to generate unseen evasion samples. The purpose is to annex the generated data with the original train set for adversarial training to improve the detection performance of machine learning (ML) classifiers. The quality of generated adversarial samples relies on the adequacy of training data samples. However, in low data regimes like medical diagnostic imaging and cybersecurity, the anomaly samples are scarce in number. This paper proposes a novel GAN design called Evasion Generative Adversarial Network (EVAGAN) that is more suitable for low data regime problems that use oversampling for detection improvement of ML classifiers. EVAGAN not only can generate evasion samples, but its discriminator can act as an evasion-aware classifier. We have considered Auxiliary Classifier GAN (ACGAN) as a benchmark to evaluate the performance of EVAGAN on cybersecurity (ISCX-2014, CIC-2017 and CIC2018) botnet and computer vision (MNIST) datasets. We demonstrate that EVAGAN outperforms ACGAN for unbalanced datasets with respect to detection performance, training stability and time complexity. EVAGAN's generator quickly learns to generate the low sample class and hardens its discriminator simultaneously. In contrast to ML classifiers that require security hardening after being adversarially trained by GAN-generated data, EVAGAN renders it needless. The experimental analysis proves that EVAGAN is an efficient evasion hardened model for low data regimes for the selected cybersecurity and computer vision datasets. Code will be available at HTTPS://www.github.com/rhr407/EVAGAN.
\end{abstract}

\begin{IEEEImpStatement}
Artificial Intelligence (AI) applications can help improve the quality of human life. The use of AI is not only limited to medical anomaly detection and drug discovery but can be leveraged in computer networks to keep people safe from malicious activities on the Internet. However, the AI-based models can be biased towards the majority class of data on which they are trained due to data imbalance. Anomaly data samples are always scarce as compared to normal data samples. So this is an open research problem to solve. Our work is an effort to improve the AI-based methods in detection performance, stability and time complexity. Using the proposed technique, we can train our AI model using fewer anomaly samples, improving the cost-efficiency compared to state-of-the-art in anomaly detection.
\end{IEEEImpStatement}


\section{Introduction}
\label{sec: intro}

Low data regimes are found in many real-life applications in which researchers face data scarcity problems \cite{moret2020generative}. The data scarcity pertains to the situation where one class is abundant in data samples (especially normal behaviour) while the anomaly samples are rare and challenging to gather \cite{vu2020handling}. The data scarcity can also be described as a data imbalance problem potentially resulting in decision bias in machine learning (ML) classifiers. The network traffic datasets are one of the prime examples of data imbalance problems. Since the ML intrusion detection systems are data-hungry probabilistic models, having more data can improve their performance \cite{shahriar2020g}. The real attacks can be emulated with dedicated machines in a lab environment using open-source operating systems like Kali Linux \cite{sharafaldin2018toward, lashkari2017characterization}. However, there can be two main disadvantages of emulating real attacks: First, real data gathering can be expensive, involving multiple hardware resources like multiple computers and network switches \cite{koroniotis2019towards}. Second, the emulated attacks may not accurately represent a real attack scenario. A cost-effective way of gathering the attacks' data is synthetic generation using AI generative models \cite{randhawa2021security}. 

Synthetic data generation is also termed data oversampling. Using generative adversarial networks (GANs) as synthetic oversamplers has been a voguish research endeavour for low data regimes \cite{shahriar2020g, sushko2021one}. Various researchers have demonstrated that GANs are more effective as compared to other synthetic oversamplers like SMOTE \cite{vu2020handling, engelmann2020conditional, randhawa2021security, kovacs2019smote}. It is found in numerous studies that due to the adversarial factor, GANs can better estimate the target probability distribution \cite{vu2020handling, engelmann2020conditional, creditcard}. In a simple/vanilla GAN, two different neural networks generator ($\mathcal{G}$) and discriminator ($\mathcal{D}$) work antagonistically to learn from each other's experience to converge to Nash equilibrium \cite{goodfellow2014generative}. As an oversampler, after being trained to a certain number of epochs, $\mathcal{G}$ is used to generate additional data. Depending on how well a GAN learned the input data probability distribution, the close resembling data is annexed to the original train set. This process is called data augmentation (DA), which many researchers have demonstrated to be effective in improving the detection performance of ML classifiers \cite{antoniou2017data, merino2019expansion, usama2019generative, lin2018idsgan, zhang2019deep}.

Since AI-based systems are prone to adversarial evasion attacks, it is imperative to harden the ML classifiers against adversarial evasions. Black box attackers can use GANs to generate evasion samples \cite{usama2019generative, lin2018idsgan, yan2019automatically}. Therefore, employing GANs can be an effective technique to proactively design an adversarial aware classifier resulting from DA. Although DA is effective in helping the ML classifiers recognise the perturbed data samples, $\mathcal{D}$ of a GAN can be extended to act as a multiclass classifier so that it can be used as an anomaly detector \cite{ yin2019enhancing, yin2018enhanc, lin2018idsgan}. In this way, we do not need to use DA as the $\mathcal{D}$ is trained simultaneously with $\mathcal{G}$. Auxiliary Classifier GAN (ACGAN) is an example of such a GAN in which the $\mathcal{D}$ not only differentiates between fake and real samples but also can be used as a multiclass classifier \cite{vu2020handling, odena2017conditional}. The advantage of extending the $\mathcal{D}$ in ACGAN is to improve training stability, and quality of generated samples \cite{odena2017conditional}.
In this work, with the help of experimentation, we have demonstrated that ACGAN does not perform well in highly unbalanced datasets. So we propose a novel GAN based on ACGAN called EVAGAN that outperforms ACGAN in terms of detection performance, stability in training and time complexity.

We summarise the main contributions of this paper in the following aspects: 
\begin{enumerate}

\item We propose a novel GAN model to design an evasion-aware discriminator as a sophisticated botnet detector.
\item We demonstrate by experiments that the existing use of ACGAN to design a sophisticated classifier can fail in highly unbalanced datasets.
\item We determine that EVAGAN outperforms ACGAN in terms of performance detection, stability and time complexity for cybersecurity (CC) botnet and computer vision (CV) datasets.

\end{enumerate}

Table \ref{table: main notations} shows the main notations used in this paper. The rest of this paper has been organized as follows. Section \ref{sec: background} provides a comprehensive background of vanilla GANs, data oversampling, adversarial evasion and ACGAN, section \ref{sec: evagan} presents the details of the proposed model, section \ref{sec: ImplementationDetails} gives a description of implementation details, section \ref{sec: results} demonstrates the results, section \ref{sec: discussion} provides an analysis of the results and section \ref{sec: conclusion} concludes the paper. 

\section{Background} 
\label{sec: background}

\begin{table}[tb!]

\centering
\caption{Main notations}

\label{table: main notations}

   \begin{tabular}{lcc}

      \hline
      
      \multicolumn{1}{|c}{\textbf{Notation}}&
      \multicolumn{1}{|c|}{\textbf{Definition}}\\
      
      \hline
      
      \multicolumn{1}{|c}{$\mathcal{G}$}&
      \multicolumn{1}{|c|}{Generator}\\
      
      \hline
      
      \multicolumn{1}{|c}{$\mathcal{D}$}&
      \multicolumn{1}{|c|}{Discriminator}\\
      
      \hline
      
      \multicolumn{1}{|c}{z}&
      \multicolumn{1}{|c|}{Normal distribution from noise space}\\
      
      \hline
      
      \multicolumn{1}{|c}{${z}$}&
      \multicolumn{1}{|c|}{Noise samples}\\
      
      \hline
      
      \multicolumn{1}{|c}{$p_{data}$}&
      \multicolumn{1}{|c|}{Probability distribution of real samples}\\
      
      \hline
      
      \multicolumn{1}{|c}{$p_{z}$}&
      \multicolumn{1}{|c|}{Probability distribution of noise samples}\\
      
      \hline
      
      \multicolumn{1}{|c}{$\mathcal{X}$}&
      \multicolumn{1}{|c|}{Real data distribution}\\
      
      \hline
      
      \multicolumn{1}{|c}{$\mathbb{E}$}&
      \multicolumn{1}{|c|}{Expected value}\\
      \hline

      \multicolumn{1}{|c}{$c_m$}&
      \multicolumn{1}{|c|}{Minority class labels}\\
      \hline
      
      \multicolumn{1}{|c}{$c_M$}&
      \multicolumn{1}{|c|}{Majority class labels}\\
      \hline

      \multicolumn{1}{|c}{$y_{x_{i}}$}&
      \multicolumn{1}{|c|}{Actual label of sample $x_i$ in dataset $\mathcal{X}$}\\
      \hline

   \end{tabular}

 \end{table}

\subsection{Generative Adversarial Networks (GANs)}
\label{subsec: gans}
\begin{figure}[tb!]
\centering
\includegraphics[width=8.8cm, height=5cm]{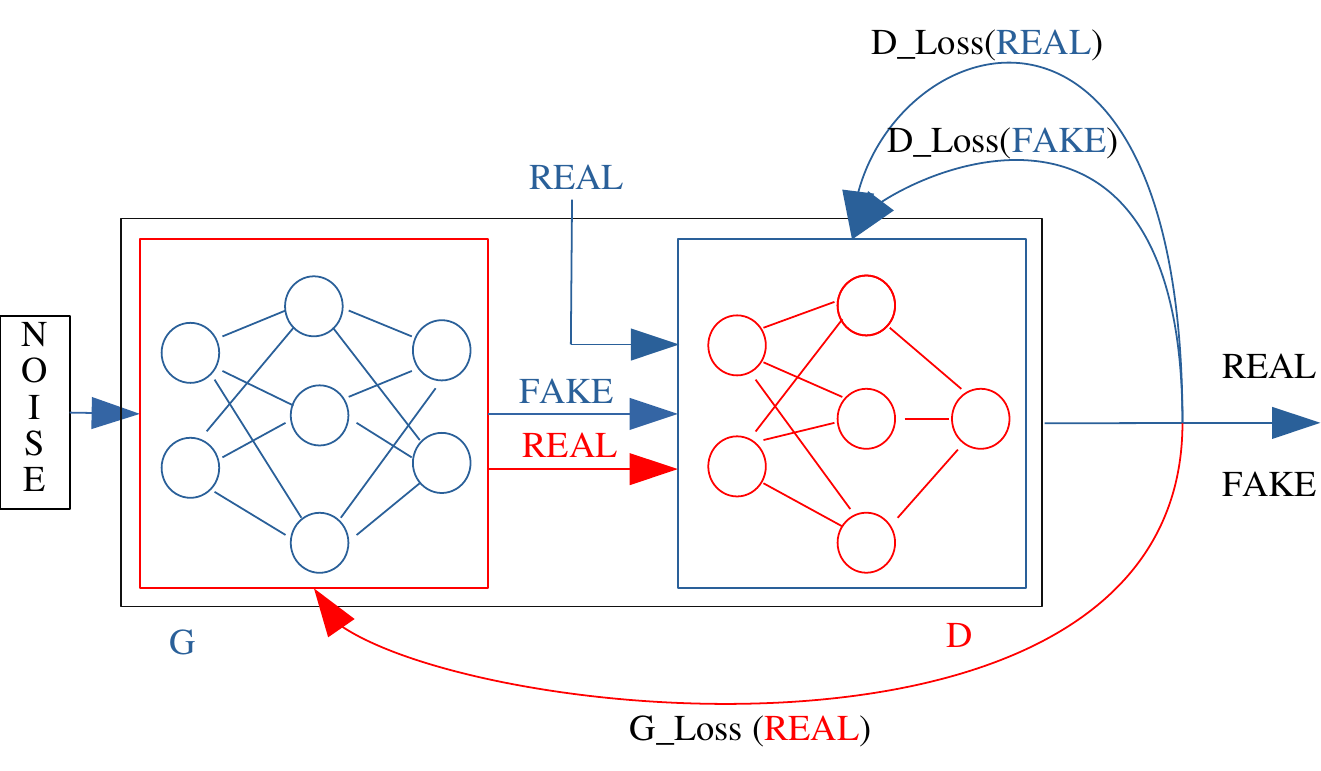}
\caption{Block Diagram of a classical GAN}
\label{fig: BlockDigram_of_GAN}
\end{figure}

A GAN combines two different neural networks, each having a unique structure. The one responsible for generating synthetic samples is called generator ($\mathcal{G}$), and the other that evaluates the generated samples is called discriminator ($\mathcal{D}$). Figure \ref{fig: BlockDigram_of_GAN} shows the block diagram of a classical/vanilla GAN. There are two consecutive steps in which a GAN is trained. In the first step, the $\mathcal{D}$ is trained on real data labelled as REAL, and the data generated by an untrained $\mathcal{G}$ is labelled as FAKE. In the next step, now that the $\mathcal{D}$ has trained already, it is tested on the fake data from $\mathcal{G}$, but this time intentionally labelled as REAL. The loss of the $\mathcal{D}$ on this falsely labelled data is fed back to the $\mathcal{G}$ which adjusts its weights in one complete batch training. There can be several batch iterations, after which one complete traversal of the dataset is complete, also known as an epoch. In the classical GAN, the generator model can be represented as $\mathcal{G}$: z$\,\to\,$ $\mathcal{X}$ where z is the normal distribution from noise space and $\mathcal{X}$ is the real data distribution. 

The discriminator $\mathcal{D}$: $\mathcal{X}$ $\,\to\,$[0,1] model is a classifier that outputs an estimate of probability between 0 and 1 to mark whether the input data is real or fake. The objective function of the combined model can be represented by Equation \ref{eq1}.

\begin{equation} \label{eq1}
\begin{aligned}
\min_\mathcal{G} \max_\mathcal{D} V(\mathcal{\mathcal{D}}, \mathcal{G})= \mathbb{E}_{x\sim p_{data}(x)}[\log \mathcal{D}(x)] + \\ \mathbb{E}_{z\sim p_z(z)}[\log(1 - \mathcal{D}(\mathcal{G}(z)))]
 \end{aligned}
\end{equation}

Here, $\mathbb{E}$ represents the expected value of the loss, and $x$ and $z$ denote the real and noise samples, respectively. At the same time, $p_{data}$ and $p_{z}$ are the probability distributions of real and noise data, respectively. The objective of a min-max game is to minimise the generator's loss in creating data resembling real data. Since the generator can not control the loss of $\mathcal{D}$ on real data, still, it can maximise the loss of $\mathcal{D}$ on generated data $\mathcal{G}(z)$. The objective function of $\mathcal{G}$ is given by Equation \ref{eq2}.

\begin{equation} \label{eq2}
\begin{aligned}
J^\mathcal{G}(\mathcal{G})= \mathbb{E}_{z\sim p_z(z)}[\log(\mathcal{D}(\mathcal{G}(z)))]
 \end{aligned}
\end{equation}

As demonstrated in the Figure \ref{fig: BlockDigram_of_GAN}, the losses of $\mathcal{D}$ on real $D\_Loss(REAL)$ and generated data $D\_Loss(FAKE)$ respectively, are fed to $\mathcal{D}$ using back-propagation. In the next step, in forward propagation, given label as REAL to the input generated samples (coming from $\mathcal{G}$), the evaluation is done by $\mathcal{D}$ and $G\_Loss(REAL)$, is fed back to $\mathcal{G}$ to update its weights. We call this step the combined model training. The combined model takes noise as input and the output of the $\mathcal{D}$ as the feedback to update the weights of the $\mathcal{G}$. This process keeps iterating till the number of epochs reaches a set value. The generator and discriminator do not learn further upon achieving the Nash equilibrium. 

\subsection{Data Oversampling \& GANs}

In low data regimes, oversampling or undersampling can help balance the datasets. However, undersampling might result in the loss of diversity. For oversampling, methods like SMOTE use nearest neighbours, and linear interpolation, which can be unsuitable for high-dimensional and complex probability distributions \cite{chawla2002smote, engelmann2020conditional}. Recent research works proposed algorithms for data oversampling. Authors in \cite{kovacs2019empirical} compared 85 different oversampling techniques and suggested the three best-performing variants as SMOTE\_IPF, ProWSyn and polynom\_fit\_SMOTE. In \cite{randhawa2021security}, authors have compared the performance of these three SMOTE variants with GANs. Through empirical results, they found that GANs outperform the three mentioned oversamplers in most of the adversarial training of ML classifiers. 

\subsection{ACGAN}

ACGAN extends a classical GAN exploiting class labels in the training process \cite{odena2017conditional}. Similar to a classical GAN, ACGAN includes two neural networks: a Generator ($\mathcal{G}$) and a Discriminator ($\mathcal{D}$). In addition to random noise samples $z$, the input of $\mathcal{G}$ includes class labels $c$. Therefore, the synthesized sample from $\mathcal{G}$ in ACGAN is $\mathcal{X}_{fake} = \mathcal{G} (c, z)$, instead of $\mathcal{X}_{fake} = \mathcal{G} (z)$. In other terms, ACGAN can generate the specified class data for which we feed labels to its $\mathcal{G}$. Simultaneously, the $\mathcal{D}$ of ACGAN works as a dual classifier for differentiating between the real/fake data and different classes of the input samples, whether coming from the real source or the $\mathcal{G}$. 

The objective function of ACGAN consists of two parts: The first is the log-likelihood $L_S$ of the correct source data, and the second is the log-likelihood $L_C$ of the real class labels. $\mathcal{D}$ is trained to maximise $L_{C} + L_{S}$ and $\mathcal{G}$ learns to maximise $L_{C} - L_{S}$. In other words, the objective of $\mathcal{D}$ is to improve the two likelihoods, while the goal of $\mathcal{G}$ is to assist $\mathcal{D}$ in improving the performance on class label discrimination. $\mathcal{G}$ will also try to suppress the log-likelihood of $\mathcal{D}$ on fake samples. The $\mathcal{D}$ outputs both a probability distribution over sources and the class labels respectively [$P (S | \mathcal{X}), P (C | \mathcal{X})] = \mathcal{D}(\mathcal{X})$ where $S$ are the sources (real/fake) and $C$ are the class labels. Equations \ref{eq3} and \ref{eq4} denote the $L_s$ and $L_c$ respectively.

\begin{equation} \label{eq3}
\begin{aligned}
L_{S} = \mathbb{E}[log P (S = real | \mathcal{X}_{real} )]\ + \\
\mathbb{E}[log P (S = fake | \mathcal{X}_{fake} )]
 \end{aligned}
\end{equation}

\begin{equation} \label{eq4}
\begin{aligned}
L_{C} = \mathbb{E}[log P (C = c| \mathcal{X}_{real} )]\ + \\
\mathbb{E}[log P (C = c | \mathcal{X}_{fake} )]
 \end{aligned}
\end{equation}

\begin{figure}[tb!]
\centering
\includegraphics[width=8.8cm, height=4.5cm]{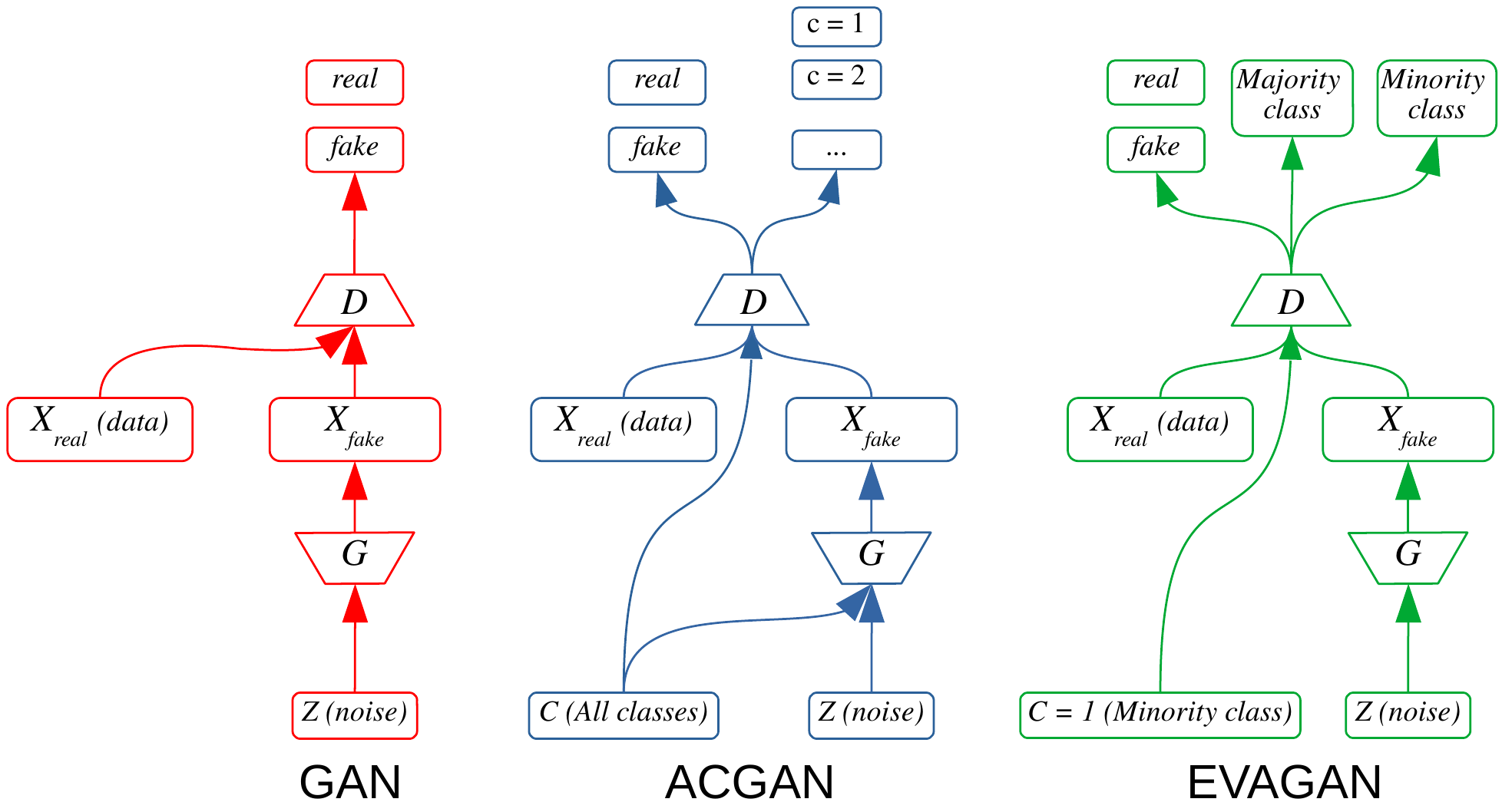}
\caption{Comparison of EVAGAN model with vanilla GAN and ACGAN}
\label{fig: GAN_vs_ACGAN_vs_EVAGAN}
\end{figure}

A careful observation of Figure \ref{fig: GAN_vs_ACGAN_vs_EVAGAN} suggests that there seems to be no tremendous difference between ACGAN and EVAGAN; however, the significance of simple modifications in the generator input, discriminator output and loss functions is discussed in more detail in section \ref{sec: evagan}. 
\subsection{Adversarial Evasion \& GANs}

The decision bias in ML classifiers can lead to the misclassification of malicious samples as normal. The attackers can exploit this intrinsic nature of ML classifiers to incarnate evasion samples, particularly in low data regimes. The adversarial evasion j* is a perturbed version of an input sample j such that the j* = j + $\eta$, where $\eta$ is a carefully crafted perturbation. When making an adversarial attack, $\eta$ could be sought and selected so that the classifier can not discriminate the j* from j \cite{papernot2017practical, apruzzese2020deep}. The researchers usually employ adversarial training to make the classifiers proactively aware of the evasion samples. However, this is not needed if we use the $\mathcal{D}$ of a GAN as a classifier to differentiate not only between the fake and real samples but also between normal and anomaly samples. The fake samples generated by the $\mathcal{G}$ are also learnt at the same time, so it is better to consider the power of $\mathcal{D}$ as an evasion-aware classifier. We do not need to use extra ML classifiers, which is a common practice in various literary works, to design such a classifier \cite{yin2018enhanc, yin2019enhancing}. 

To this end, we propose EVAGAN that provides such type of $\mathcal{D}$ and compare its performance with the $\mathcal{D}$ of ACGAN and other ML classifiers, xgboost (XGB), decision tree (DT), naive bayes (NB), random forests (RF), logistic regression (LR) and k-nearest neighbours (KNN). Following rigorous experimentation, we explore that EVAGAN's $\mathcal{D}$ not only outperforms the ML classifiers in black box testing but also gives 100\% accuracy in normal and evasion samples estimation. The details of the experimental results will be discussed in section \ref{sec: discussion}.

\section{EVAGAN}
\label{sec: evagan}

In this section, we discuss the motivation behind the design of EVAGAN, the structural explanation of its generator and discriminator, along with the objective and loss functions. 
\subsection{Motivation}
Considering the generator ($\mathcal{G}$) of ACGAN, $\mathcal{X}_{fake} = \mathcal{G} (c, z)$ where $c$ is the class label, $\mathcal{G}$ has to generate the samples of all classes. Hence the number of the samples generated by $\mathcal{G}$ may include $C = \{c_1, c_2, c_3, ... , c_n\}$ which may not be a requirement in low data regimes. Since we only need to generate a low sample class with labels $c_m$ instead of all the classes, so the generator does not need to be aware of the classification performance of $\mathcal{D}$ on majority class samples. In this way, the training time of $\mathcal{G}$ is reduced as the diversity seen by the $\mathcal{G}$ is less complex to generate a single class sample. Due to this reason, we can not only improve the performance of the $\mathcal{G}$ but also can harden the $\mathcal{D}$ simultaneously with fewer $c_m$ samples. The ratio of the different class labels can vary the performance of $\mathcal{G}$ as this is a stochastic process. However, in most cases, the normal class samples will be more than the anomaly samples. Note that EVAGAN design is dedicated to binary class problems where the samples of a minority class are scanty. For using EVAGAN for multiclass cases, each anomaly class should be considered separately from the normal class to make it a binary classification problem. However, the concept can be extended to multiclass, which we leave to future work.

\begin{figure*}[tb!]
\centering
\includegraphics[width=16cm, height=7cm]{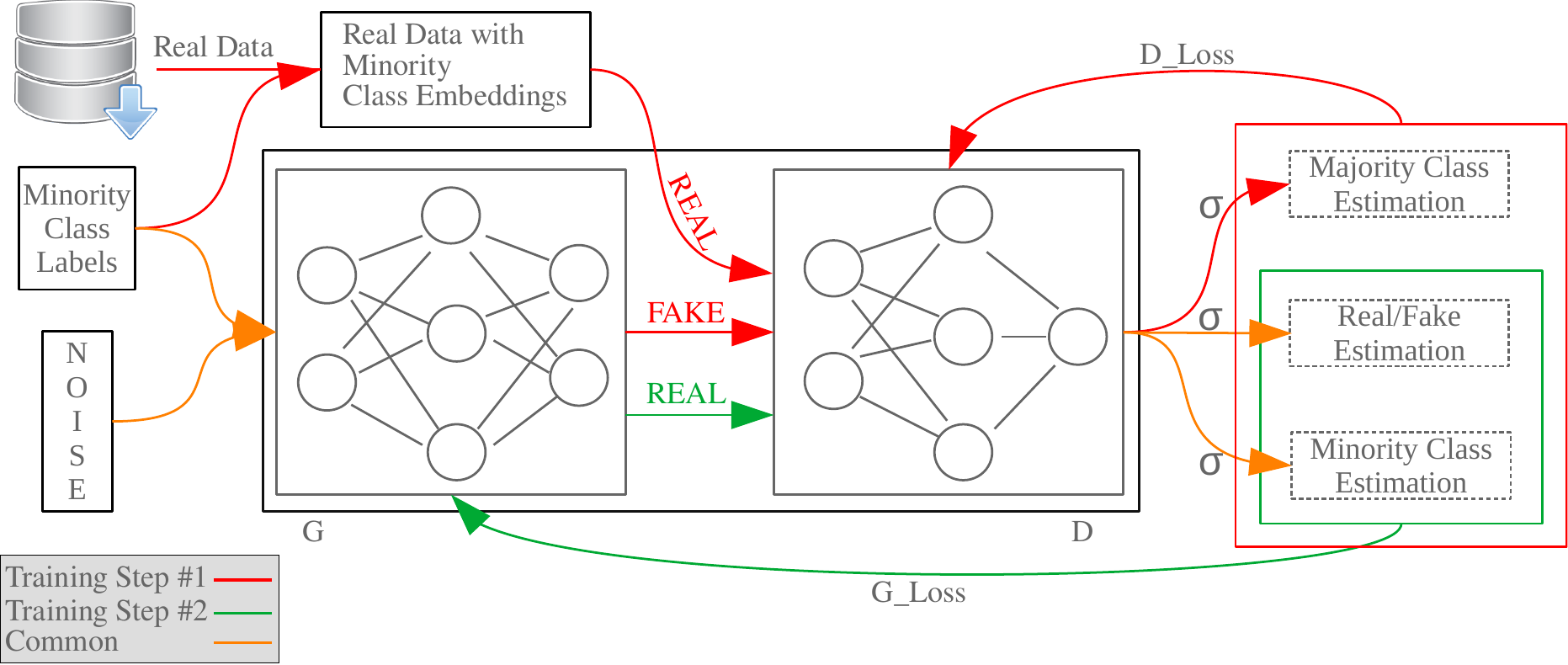}
\caption{EVAGAN Architecture}
\label{fig: EVAGAN}
\end{figure*}

\subsection{Architecture}
The design of EVAGAN is inspired by ACGAN as we want to develop a classifier model that hardens itself on the GAN-generated evasion samples. The main structure of EVAGAN consists of two neural networks; the generator $\mathcal{G}$ and the discriminator $\mathcal{D}$. In contrast to ACGAN, EVAGAN's model is limited to labels from a single class embedded with noise as the input to the generator ($\mathcal{G}$). The details of the $\mathcal{G}$ have been explained in subsection \ref{subsec: generator}. Figure \ref{fig: EVAGAN} shows the detailed architecture of EVAGAN. There are three types of colour-graded arrows shown in the figure. The red coloured arrows demonstrate the first training step in which only $\mathcal{D}$ is trained. The green arrows depict the second training step for the $\mathcal{G}$. The orange arrows show the involvement of common inputs (minority class labels and noise) and outputs (real/fake and minority class estimations) for both training steps mentioned previously. These two steps of a typical GAN training were expressed earlier in subsection \ref{subsec: gans}. The discriminator $\mathcal{D}$ of EVAGAN has three different outputs for the estimation of majority, minority and fake/real classes. Sigmoid functions have been used for the three outputs, each with binary cross-entropy (BCE) loss. The details of $\mathcal{D}$ are further expressed in subsection \ref{subsec: discriminator}. The loss functions have also been mentioned in respective subsections of the $\mathcal{G}$ and $\mathcal{D}$. 

Figure \ref{fig: EVAGAN} shows a red outlined box on the right side to illustrate the three different probability estimations as outputs from $\mathcal{D}$. These three estimations are used to compute the loss of $\mathcal{D}$ in the first step of EVAGAN training. A green outlined box, including the real/fake estimation and minority class estimation, computes the $G\_Loss$ to be fed back to the $\mathcal{G}$ in the backpropagation of the combined model training (second step of EVAGAN training). Note that the output of the $\mathcal{D}$ is distributed using three different sigmoid units to separate the probabilities of each class, i.e. the majority, sources (real/fake) and minority. The majority and minority class estimations could be combined using a single sigmoid function. However, keeping them separate has three advantages. The first is to avoid the loss of the majority class being fed back to the $\mathcal{G}$. Second, it simplifies the model with no extra training cost. Third, we can conveniently separate the predictions for the test set samples, which will be discussed in section \ref{sec: results}.

\subsection{Generator}
\label{subsec: generator}
The generator ($\mathcal{G}$) of EVAGAN only takes noise $n$ and the single class labels $c$ = 1. The labels are embedded in the input layer of the $\mathcal{G}$. The objective function of the $\mathcal{G}$ has two parts, as shown in Equations \ref{eq5} and \ref{eq6}.

\begin{equation} \label{eq5}
\begin{aligned}
I^\mathcal{G}(\mathcal{G})= \mathbb{E}_{z\sim p_z(z)}[\log(\mathcal{D}(\mathcal{G}(z)))]
 \end{aligned}
\end{equation}

\begin{equation} \label{eq6}
\begin{aligned}
J^\mathcal{G}(\mathcal{G})= \mathbb{E}_{c_m\sim y_m}[\log P(C = c_{m} | \mathcal{X}_{m_{fake}})]
 \end{aligned}
\end{equation}

Equation \ref{eq5} is the objective function of $\mathcal{G}$ similar to Equation \ref{eq2}. The goal is to minimise the log-likelihood of the fake samples being classified as fake by $\mathcal{D}$. In Equation \ref{eq6}, $J^\mathcal{G}(\mathcal{G})$ is the objective function of $\mathcal{G}$ for improving the log-likelihood of minority class samples coming from the $\mathcal{G}$ into the $\mathcal{D}$. Here, $y_m$ denotes the minority class label in the real dataset, and $P$ is the output probability from $\mathcal{D}$. Since the $\mathcal{G}$ only needs to generate $c_m$ samples so it should only receive the loss of $\mathcal{D}$ on the estimation of minority class and the sources, i.e. the samples being real or fake. The objective function of $\mathcal{G}$ is to maximise the $\mathcal{D}$ loss on the fake source. At the same time, it will assist in minimising the $\mathcal{D}$ loss on $c_m$ samples. Equation \ref{eq7} shows the objective function of $\mathcal{G}$.
\begin{equation} \label{eq7}
\begin{aligned}
L^\mathcal{G}(\mathcal{G}) = J^\mathcal{G}(\mathcal{G}) - I^\mathcal{G}(\mathcal{G})
 \end{aligned}
\end{equation}

The cross-entropy (CE) loss of two different probability distributions $p(x)$ and $q(x)$ can be denoted using Equation \ref{eq8}, where x denotes the samples belonging to the $\mathcal{X}$ dataset.

\begin{equation} \label{eq8}
\begin{aligned}
CE(p,q) = - \sum\limits_{x\epsilon X} p(x) \log q(x)
 \end{aligned}
\end{equation}

Let $y_{x_{i}}$ be the actual label of sample $x_i$ in dataset $\mathcal{X}$, $P(S = fake | \mathcal{X}_{m_{fake}})$ be the predicted probability distribution of generated samples being fake and $P(C = c_m | \mathcal{X}_{m_{fake}})$ be the predicted probability distribution from $\mathcal{D}$ for minority class labels $c_m$, then the loss function of $\mathcal{G}$ for $N$ samples will be given by the Equation \ref{eq9}.

\begin{equation} \label{eq9}
\begin{aligned}
G\_Loss = - \frac{1}{N}\sum\limits_{i = 1}^{N} [
y^{fake}_{x_{i}} (\log P(S = fake | \mathcal{X}_{m_{fake}})) + \\
y^{c_m}_{x_{i}} (1 - \log P(C = c_m | \mathcal{X}_{m_{fake}})) ]
 \end{aligned}
\end{equation}

In Equation \ref{eq9}, $y^{fake}_{x_{i}}$ and $y^{c_m}_{x_{i}}$ are the actual labels for fake and minority classes respectively. According to Equation \ref{eq9}, the goal of $\mathcal{G}$ is to minimize the $G\_Loss$, so it tends to reduce the correct estimation of $\mathcal{D}$ on fake samples by suppressing the term $\log P(S = fake | \mathcal{X}_{m_{fake}})$. For the second objective, it will try to increase the value of $\log P(C = c_m | \mathcal{X}_{m_{fake}})$ so that the second term in the equation can also be suppressed in value.

\subsection{Discriminator}
\label{subsec: discriminator}
For the $\mathcal{D}$ model of EVAGAN, we have separated the majority, and minority class estimations using two different sigmoid ($\sigma$) functions as demonstrated in Figure \ref{fig: EVAGAN}. The benefit of separating the majority and minority class estimations is that we can feedback only minority class estimation to the $\mathcal{G}$. The other advantage of this structure is that we can separately calculate the estimation of both classes on test datasets to compare it with the ACGAN model later done in section \ref{sec: results}. The objective function of $\mathcal{D}$ has three parts as given by the Equations \ref{eq10}, \ref{eq11} and \ref{eq12}. For the minority class terminologies, we use '$m$', and for the majority class, we use '$M$' in the following equations.

\begin{equation} \label{eq10}
\begin{aligned}
L_M = \mathbb{E}_{c_M\sim y_{M_{real}}} [\log P(C=c_M|\mathcal{X}_{M_{real}})]
 \end{aligned}
\end{equation}

\begin{equation} \label{eq11}
\begin{aligned}
L_{S_m} = \mathbb{E}_{y_{m_{real}}} [log P (S = real | \mathcal{X}_{m_{real}} )]\ + \\
\mathbb{E}_{y_{m_{fake}}} [log P (S = fake | \mathcal{X}_{m_{fake}} )]
 \end{aligned}
\end{equation}

\begin{equation} \label{eq12}
\begin{aligned}
L_m = \mathbb{E}_{c_m\sim y_{m_{real}}} [\log P(C=c_m|\mathcal{X}_{m_{real}})]\ + \\ 
\mathbb{E}_{c_m\sim y_{m_{fake}}} [ \log P (C=c_m | \mathcal{X}_{m_{fake}} )]
 \end{aligned}
\end{equation}

The first goal of the $\mathcal{D}$ is to correctly estimate the majority class distribution from the real samples only as $\mathcal{G}$ does not generate the majority class samples. Equation \ref{eq10} denotes the log-likelihood for the real majority class samples. Equation \ref{eq11} represents the source log-likelihood for the real and fake minority class samples. Equation \ref{eq12} summarises the real and fake log-likelihoods from the $\mathcal{D}$ for minority class samples. Hence, the objective function of the $\mathcal{D}$ can be represented as the sum of the three log-likelihoods to be maximised by the $\mathcal{D}$ as given by Equation \ref{eq13} 

\begin{equation} \label{eq13}
\begin{aligned}
L^\mathcal{D}(\mathcal{D}) = L_M + L_{{S}_m} +L_m
 \end{aligned}
\end{equation}

The loss function of $\mathcal{D}$ can be derived using Equation \ref{eq5} and \ref{eq6}, given by Equation \ref{eq14}.

\begin{equation} \label{eq14}
\begin{aligned}
D\_Loss = - \frac{1}{N}\sum\limits_{i = 1}^{N} [
y^{c_M}_{x_{i}} (\log P(S = c_M | \mathcal{X}_{M_{real}})) + \\
y^{real}_{x_{i}} (\log P(S = real | \mathcal{X}_{m_{real}})) + \\
(1 - y^{real}_{x_{i}}) (1 - \log P(S = real | \mathcal{X}_{m_{real}})) + \\
y^{c_{m_{real}}}_{x_{i}} (\log P(C = c_m | \mathcal{X}_{m_{real}})) + \\
(1 - y^{c_{m_{real}}}_{x_{i}}) (1 - \log P(C = c_m | \mathcal{X}_{m_{real}})) ]
 \end{aligned}
\end{equation}

In Equation \ref{eq14}, the loss of $\mathcal{D}$ has been derived from three different binary cross-entropy losses for majority class, sources and minority class estimations. Note that we have ignored the loss on $c_M$ for being fake because no majority class samples are being generated by the $\mathcal{G}$.


\section{Implementation Details}
\label{sec: ImplementationDetails}
\subsection{Experimental Setup}
The experiments were performed on a GPU workstation, AMD Ryzen threadripper 1950x with a 16-core processor and GeForce GTC 1070 Ti (8GB) graphics card, running ubuntu 20.04. Keras, TensorFlow, Sklearn and Numpy libraries were used in the Jupyter notebook and visual studio code (VSCode). The source code of EVAGAN has been provided on GitHub under MIT license\footnote{https: //github.com/rhr407/EVAGAN}.

\subsection{Data Preparation}
For experimentation, we have used CC botnet and CV MNIST datasets. The quantitative analysis of EVAGAN was performed on CC datasets. We have followed the work done by the authors in \cite{randhawa2021security} for dataset selection of botnet. We have used three datasets, ISCX-2014, CIC-2017 and CIC-2018, from the Canadian Institute of Cybersecurity (CIC). The features were extracted using a utility called CICFlowMeter-v4 provided by the CIC. We have inherited the same feature set as mentioned in \cite{randhawa2021security}. The reader may refer to this article for more details on the feature set used for the three datasets. The number of samples of benign vs botnet is mentioned in Table \ref{table: number of samples in datasets}. The qualitative analysis was performed using visual inspection. For this purpose, we used the MNIST handwriting digits dataset.

\subsection{CC Datasets}
Following is the detail of CC datasets and the botnet samples used in this work. This subsection also includes the preprocessing methodology for the selected datasets.
\subsubsection{ISCX-2014 Dataset}
The ISCX-2014 dataset \cite{beigi2014towards} is a combination of three publicly available datasets ISOT \cite{zhao2013botnet}, ISCX 2012 IDS \cite{shiravi2012toward} and CTU-13 \cite{garcia2013malware}. As per the ISCX website's details, it complies with generality, realism, and representativeness. The generality represents the richness of diversity of botnet behaviour. Realism can be defined as the closeness with the actual traffic captured, and representativeness is the ability to reflect the real environment, which a botnet detector would need in deployment. Only the Virut botnet was selected for this work because it had fewer samples than other botnets except for Zeus, which had insufficiently low samples. The labels of SMTP or NSIS were not available on the website\footnote{https: //www.unb.ca/cic/datasets/botnet.html}. Hence, we used a subset with all the normal traffic flows and Virut samples. In this way, we could use this dataset as a good example of an unbalanced set. The distribution of the normal and Virut samples has been shown in Table \ref{table: number of samples in datasets}.

\subsubsection{CIC-IDS2017 Dataset}
The botnet chosen for the CIC-IDS2017 was Ares. For this bot, the traffic was collected on Friday, July 7, 2017, from 10: 02 AM to 11: 02 AM in the CIC facility. The dataset is available on the CIC website\footnote{https: //www.unb.ca/cic/datasets/ids-2017.html}. Similar to ISCX-2014, a subset of this dataset using all the normal flows with the selected botnets was created. The ratio of the number of samples has been mentioned in Table \ref{table: number of samples in datasets}.
 
\subsubsection{CIC-IDS2018 Dataset}
To create another subset of an unbalanced dataset for analysis, we used CIC-IDS2018. This dataset included samples for Ares and Zeus botnets. We created a subset of all the normal and 2560 botnet traffic flows to generate another unbalanced dataset.

\begin{table}[ht!]

\centering
\caption{Distribution of normal and botnet samples in cybersecurity botnet datasets}

\label{table: number of samples in datasets}

   \begin{tabular}{lcccc}

      \hline

      \multicolumn{1}{|c}{\textbf{Dataset}}&
      \multicolumn{1}{|c}{\textbf{Normal}}&
      \multicolumn{1}{|c}{\textbf{Real\_bots}}&
      \multicolumn{1}{|c|}{\textbf{Total Samples}}\\
      
      \hline
      
      \multicolumn{1}{|c}{ISCX-2014}&
      \multicolumn{1}{|c}{246929}&
      \multicolumn{1}{|c}{Virut: 1748}&
      \multicolumn{1}{|c|}{248677}\\
      
      \hline
      
      \multicolumn{1}{|c}{CIC-IDS2017}&
      \multicolumn{1}{|c}{70374}&
      \multicolumn{1}{|c}{Ares: 1956}&
      \multicolumn{1}{|c|}{72330}\\
      
      \hline
      
      \multicolumn{1}{|c}{CIC-IDS2018}&
      \multicolumn{1}{|c}{390961}&
      \multicolumn{1}{|c}{Ares/Zeus: 2560}&
      \multicolumn{1}{|c|}{393521}\\
      
      \hline
    
   \end{tabular}

 \end{table}

\subsubsection{Feature Selection}
The quality of a botnet dataset determines the performance of the botnet detectors in general and the number of distinct features in particular. A reduced feature set may not perform a stronger classification as compared to an enhanced set of non-redundant features \cite{randhawa2021security}. In \cite{beigi2014towards}, the authors summarised the most important network flow features that could be helpful in botnet detection. We have used almost all of these features, which were mentioned in \cite{sharma2018machine} as well. The CICFlowMeter-v4 utility was used to extract 80 flow and time-based features\footnote{https: //www.unb.ca/cic/datasets/ids-2018.html} from their $.pcap$ files. This utility can be advantageous for extracting the mentioned features for any input $.pcap$ file.
\subsubsection{Preprocessing}
\label{subsec: preprocessing}
The ISCX-2014 dataset has not been labelled to be used in ML-based experiments. We used the information provided on the CIC website for IPs associated with the particular botnets to label the dataset. After labelling, we performed preprocessing; All the high and low skewed values were removed to suppress outliers. The columns with NaN, Inf and zero standard deviation were removed. Finally, the dataset was scaled to the [0,1] range to use rectified linear unit (ReLU) activation function in the GAN model for data generation. The CIC-IDS2017 and CIC-IDS2018 were already labelled, so we only did preprocess for these two datasets after extracting the unbalanced subsets. Our experiments used 70\% of the cybersecurity subsets as training sets, and the rest of the 30\% was used for testing the models and ML classifiers. 
\subsection{CV Dataset}
\subsubsection{MNIST Dataset}

The MNIST dataset is a simplified collection of handwritten digits ranging from 0 to 9 for training and testing various ML algorithms \cite{lecun1998mnist}. The purpose of using this dataset was to evaluate the performance of EVAGAN against ACGAN in terms of the visual quality of the images generated in balanced and unbalanced scenarios. 

\subsection{Model Comparison of EVAGAN with ACGAN}
For comparison, we constructed four different variants of GANs, respectively ACGAN\_CC, EVAGAN\_CC, ACGAN\_CV, and EVAGAN\_CV. ACGAN\_CC and EVAGAN\_CC were trained and tested on CC datasets, and ACGAN\_CV and EVAGAN\_CV used CV datasets. The implementation details of each version in terms of hyperparameters can be found in Table \ref{table: GANs_hyper}. 

\subsubsection{ACGAN\_CC \& EVAGAN\_CC}
The structure of ACGAN\_CC and EVAGAN\_CC was made up of densely connected feed-forward neural network (FFNN) for both $\mathcal{G}$ and $\mathcal{D}$. The activation functions in hidden layers for both GANs were rectified linear units (ReLU). The hidden layers were regularized using batch normalization, and the optimizer type was Adam with binary cross-entropy (BCE). The difference between ACGAN\_CC and EVAGAN\_CC is in the output layers of $\mathcal{D}$. The $\mathcal{D}$ of ACGAN\_CC outputs two neurons, one for the source probability and the other for the class probability for two classes (normal and botnet). The activation function is sigmoid for both outputs. The output layer structure of EVAGAN\_CC has three neurons, one for the normal class, the second for the source probability, and the third for the botnet class (minority class). Each of the three outputs leverages the sigmoid as the activation function.
\subsubsection{ACGAN\_CV \& EVAGAN\_CV}
The CV-based GAN architecture is different as it deals with image data compared to tabular data in CC. We need to use the convolutional neural network (CNN) instead of FFNN with other layers specific for image generation or detection. The output layer of $\mathcal{D}$ is similar to CC-based GAN implementations, except the Adam optimizer's loss function has BCE for source estimations and sparse categorical cross-entropy (SCCE) for class labels. Here, BCE could have been used; however, minimal changes to the code were made to maintain the integrity of the original ACGAN. However, in ACGAN\_CC, we have used BCE as we converted the CNN-based code to FFNN ourselves. In this way, we could keep ACGAN\_CC and EVAGAN\_CC as similar as possible for a fair comparison. 

\begin{table*}[tb!]

\centering
\caption{CC and CV GAN Models}
\label{table: GANs_hyper}
\centering
\begin{tabu} to 1\textwidth{|m{2.8cm}|m{3cm}|m{3cm}|m{3cm}|m{3cm}|} %

\hline
 
  \centering \textbf{Parameter} &
  \centering \textbf{ACGAN\_CC} &
  \centering \textbf{EVAGAN\_CC} & 
  \centering \textbf{ACGAN\_CV} & 
  \centering \textbf{EVAGAN\_CV} \\ 
  \hline
 
   \centering  Network Type & 
      \multicolumn{2}{c|}{FFNN} & 
      \multicolumn{2}{c|}{CNN}\\

  \hline

  \centering  Number of Layers  &
  \multicolumn{2}{c|}{\centering $\mathcal{G}$: 5, $\mathcal{D}$: 5} & \multicolumn{2}{c|}{\centering $\mathcal{G}$: 2, $\mathcal{D}$: 3}
  
  \\

  \hline

\centering  Activations  &
  \multicolumn{2}{c|}{$\mathcal{G}$: ReLU, $\mathcal{D}$: LeakyReLU (output: sigmoid) } &
  \multicolumn{2}{c|}{$\mathcal{G}$: ReLU (output: tanh), $\mathcal{D}$: ReLU (output: sigmoid, softmax) }
  \\
    \hline
  
  \centering Batch Size ($b$) &
  \multicolumn{4}{c|}{256}\\

\hline 

  
  \centering Neurons in input layer  &
  \multicolumn{4}{c|}{$\mathcal{G}$: latent dimension, class label vector size, $\mathcal{D}$: feature size}\\
\hline

\centering  Neurons in layer 1   &
\multicolumn{2}{c|}{$\mathcal{G}$ : 32, $\mathcal{D}$ : 128}&
$\mathcal{G}$ : 128, $\mathcal{D}$ : 32 & $\mathcal{G}$ : 128, $\mathcal{D}$ : 32  \\
  \hline
  
\centering  Neurons in layer 2   &
\multicolumn{2}{c|}{$\mathcal{G}$ : 64, $\mathcal{D}$ : 64}&
\multicolumn{2}{c|}{$\mathcal{G}$ : 64, $\mathcal{D}$ : 64} \\
  \hline

  \centering  Neurons in layer 3   &
\multicolumn{2}{c|}{$\mathcal{G}$ : 128, $\mathcal{D}$ : 32} &
\multicolumn{2}{c|}{$\mathcal{D}$ : 128}\\
  \hline
  

  \centering  Neurons in output layer   &
  \centering $\mathcal{G}$ : feature size, $\mathcal{D}$ : 2  & 
  \centering $\mathcal{G}$ : feature size, $\mathcal{D}$ : 3  
  &
  
  \centering $\mathcal{G}$ : feature size, $\mathcal{D}$ : 2  & 
  \centering $\mathcal{G}$ : feature size, $\mathcal{D}$ : 3 \\
  \hline

     \centering Layer Regularization &
  \multicolumn{4}{c|}{$\mathcal{G, D}$: $BatchNorm$ } \\

  \hline

  \centering Optimizer   &
  
\multicolumn{4}{c|}{Adam (beta\_1=0.0002, beta\_2=0.5)}\\

  \hline
  
 \centering  Loss Function  &
 \multicolumn{2}{c|}{BCE}  &
 \multicolumn{2}{c|}{BCE, SCCE} \\

  \hline
  
  \centering Learning\ Rate   &
  \multicolumn{4}{c|}{5e-4}\\ 
    \hline

    \centering Epochs   &
  \multicolumn{4}{c|}{150}\\ 

  \hline

 \end{tabu}
 \end{table*}

\section{Results}
\label{sec: results}
This section shows the results of the GAN implementations around two types of datasets: CC and CV GANs.

\subsection{CC GANs}
The results for quantitative analysis of the $\mathcal{D}$'s performance on generated samples validity (GEN\_VALIDITY), fake/generated botnet samples evasion (FAKE\_BOT\_EVA), real normal/majority class estimation (REAL\_NORMAL\_EST) and real botnet/minority class evasion (REAL\_BOT\_EVA) have been demonstrated in Figure \ref{fig: estimations_cc}. The ML classifier results have also been shown in this figure for the three CC datasets for comparison. Equations from \ref{eq15}-\ref{eq18} represent the mathematical expressions for these performance indicators. We have used Keras $model.predict$ function to compute the values where the $model$ is $\mathcal{D}$ as our prime objective is to devise an intelligent evasion aware classifier. Following is a brief detail of each evaluation parameter. 

\subsubsection{GEN\_VALIDITY}
In Equation \ref{eq15}, $\hat {\mathcal{G}}(z,c_m)[0]$ denotes the predicted value for the source being fake or real. The Keras $model.predict$ function outputs an array, so the average of the first elements in the array will be the source validity of the generated samples after every epoch. The more this value is close to '1', the more it will be regarded as real. 

\begin{equation} \label{eq15}
\begin{aligned}
GEN\_VALIDITY = \frac{\sum [
\hat {\mathcal{G}}(z,c_m)[0] ]}{N}
 \end{aligned}
\end{equation}

\subsubsection{FAKE\_BOT\_EVA}
In Equation \ref{eq16}, $\hat {\mathcal{G}}(z,c\_m)[1]$ represents the probability estimation of generated minority/botnet class samples. Since the label for minority/botnet class is '0' so ideally, we expect the model to output a value close to '0'. We represent this estimation as the evasion of the generated samples. So the more this value is close to '0', the less evasion will be. Note that this is the second value in the sum of the $model.predict$ function output. 

\begin{equation} \label{eq16}
\begin{aligned}
FAKE\_BOT\_EVA =\frac{\sum [
\hat {\mathcal{G}}(z,c_m)[1] ]}{N}
 \end{aligned}
\end{equation}

\subsubsection{REAL\_NORMAL\_EST}
In Equation \ref{eq17}, $\hat {\mathcal{X}}_{normal_{test}}[2]$ represents the probability estimation of majority/normal class samples. Since the majority/normal class label is '1', ideally, we expect the model to output a value close to '1'. Note that this is the third value in the sum of the $model.predict$ function output for the normal samples from the test set. 

\begin{equation} \label{eq17}
\begin{aligned}
REAL\_NORMAL\_EST = \frac{\sum {[
\hat {\mathcal{X}}_{normal_{test}}}[2] ]}{N}
 \end{aligned}
\end{equation}

\subsubsection{REAL\_BOT\_EVA}
In Equation \ref{eq18}, $\hat {\mathcal{X}}_{botnet_{test}}[1]$ represents the probability estimation of the real minority/botnet class samples. Our expectation from the model is to output the value close to '0', similar to FAKE\_BOT\_EVA. This is the second value in the sum of the $model.predict$ function output for the botnet samples from the test set. 

\begin{equation} \label{eq18}
\begin{aligned}
REAL\_BOT\_EVA =\frac{\sum {[
\hat {\mathcal{X}}_{botnet_{test}}}[1]]}{N}
 \end{aligned}
\end{equation}

\subsubsection{Losses}
\begin{figure}[tb!]
\centering
\includegraphics[width=8.8cm, height=6cm]{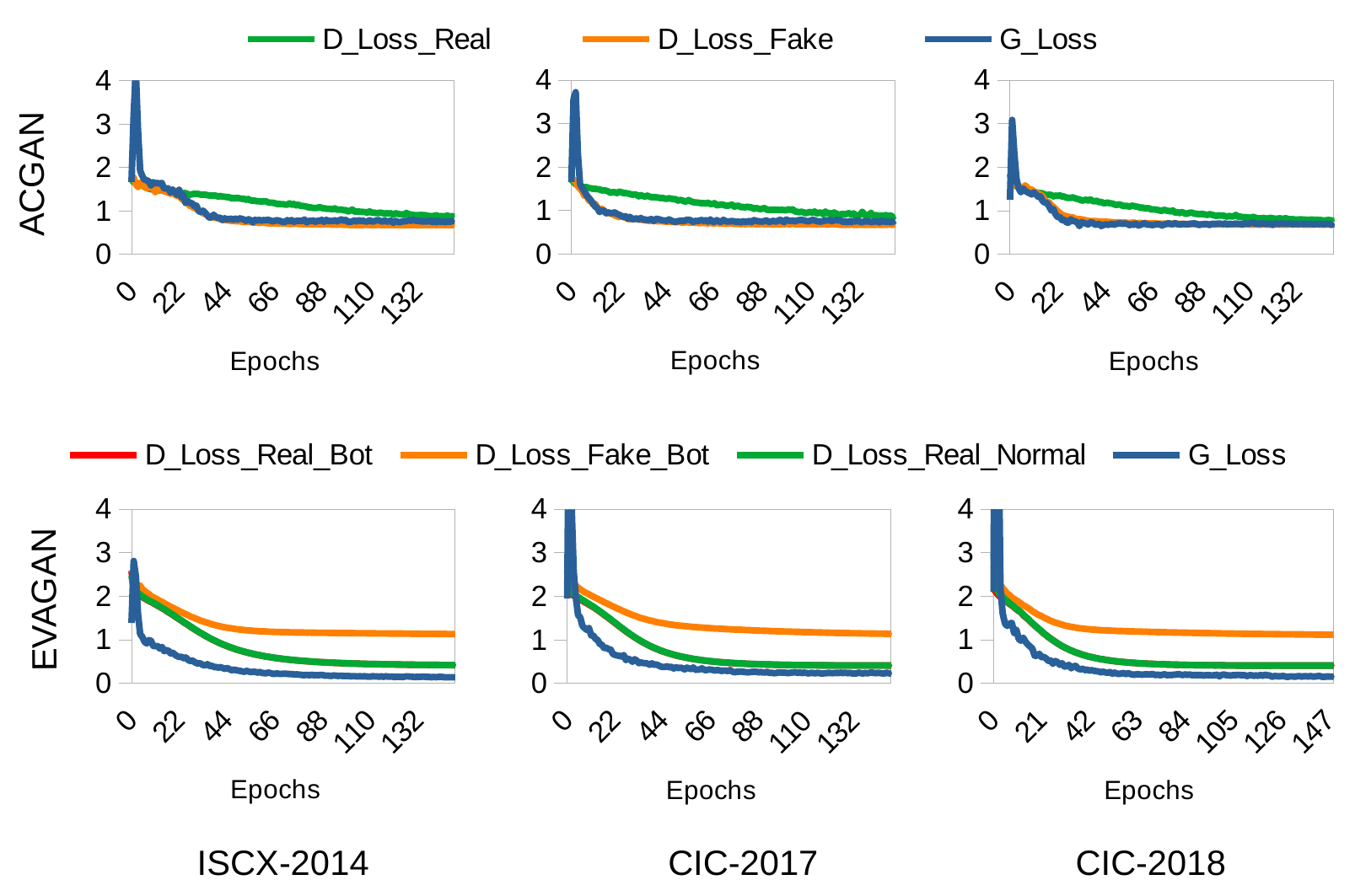}
\caption{CC GANs Losses: The training losses for ACGAN and EVAGAN on three different CC datasets}
\label{fig: losses_cc}
\end{figure}

The losses of $\mathcal{D}$ for real and fake minority classes and majority/normal class and the loss of $\mathcal{G}$ have been demonstrated in Figure \ref{fig: losses_cc} for both ACGAN and EVAGAN. 

\begin{figure}[tb!]
\centering
\includegraphics[width=8.8cm, height=8.1cm]{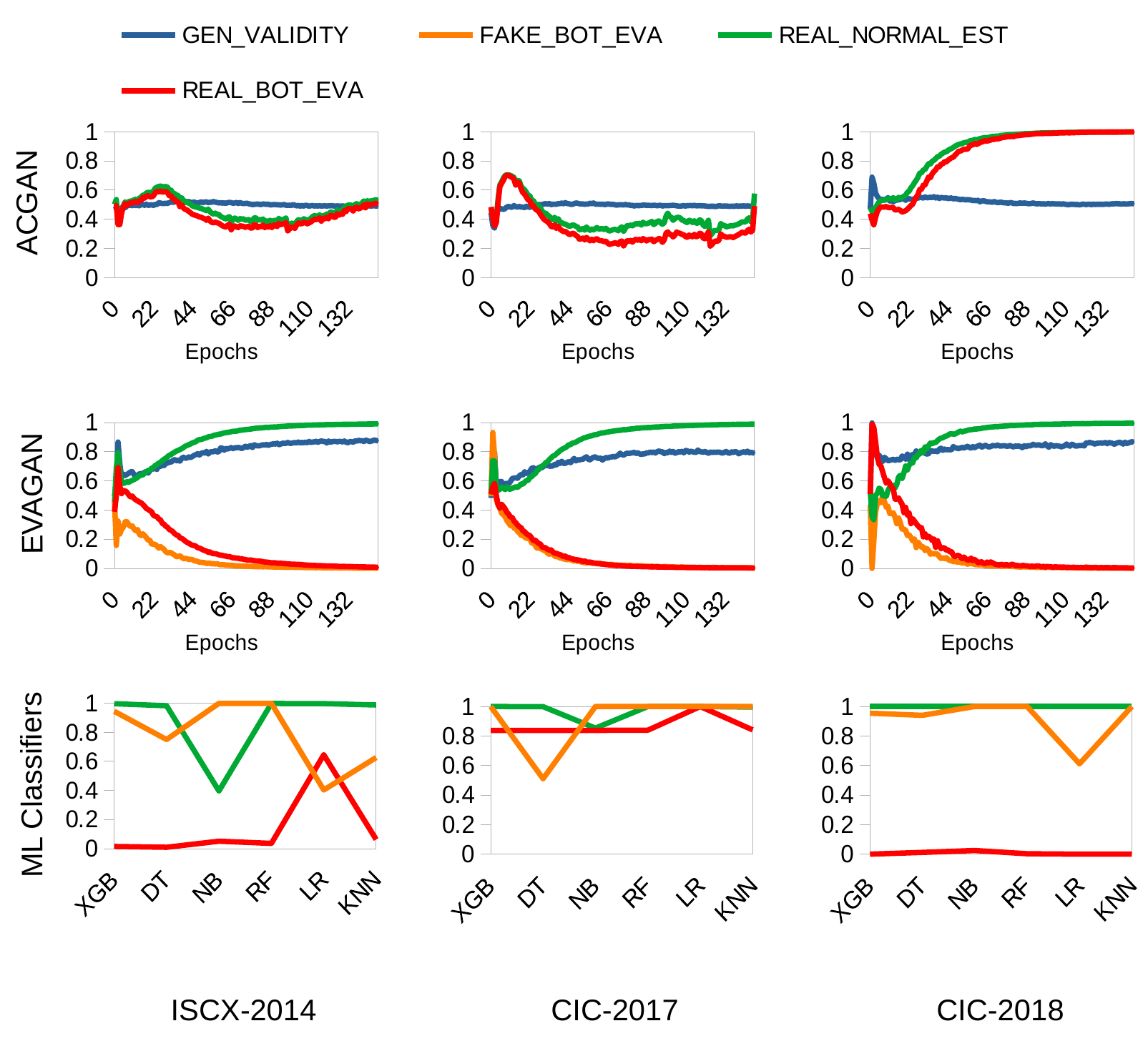}
\caption{CC Estimations: The estimations on test data and data generated by the relative GANs along with the results of six different ML-classifiers}
\label{fig: estimations_cc}
\end{figure}

\subsection{CV GANs}
\begin{figure*}[tb!]
\centering
\includegraphics[width=12.8cm, height=6cm]{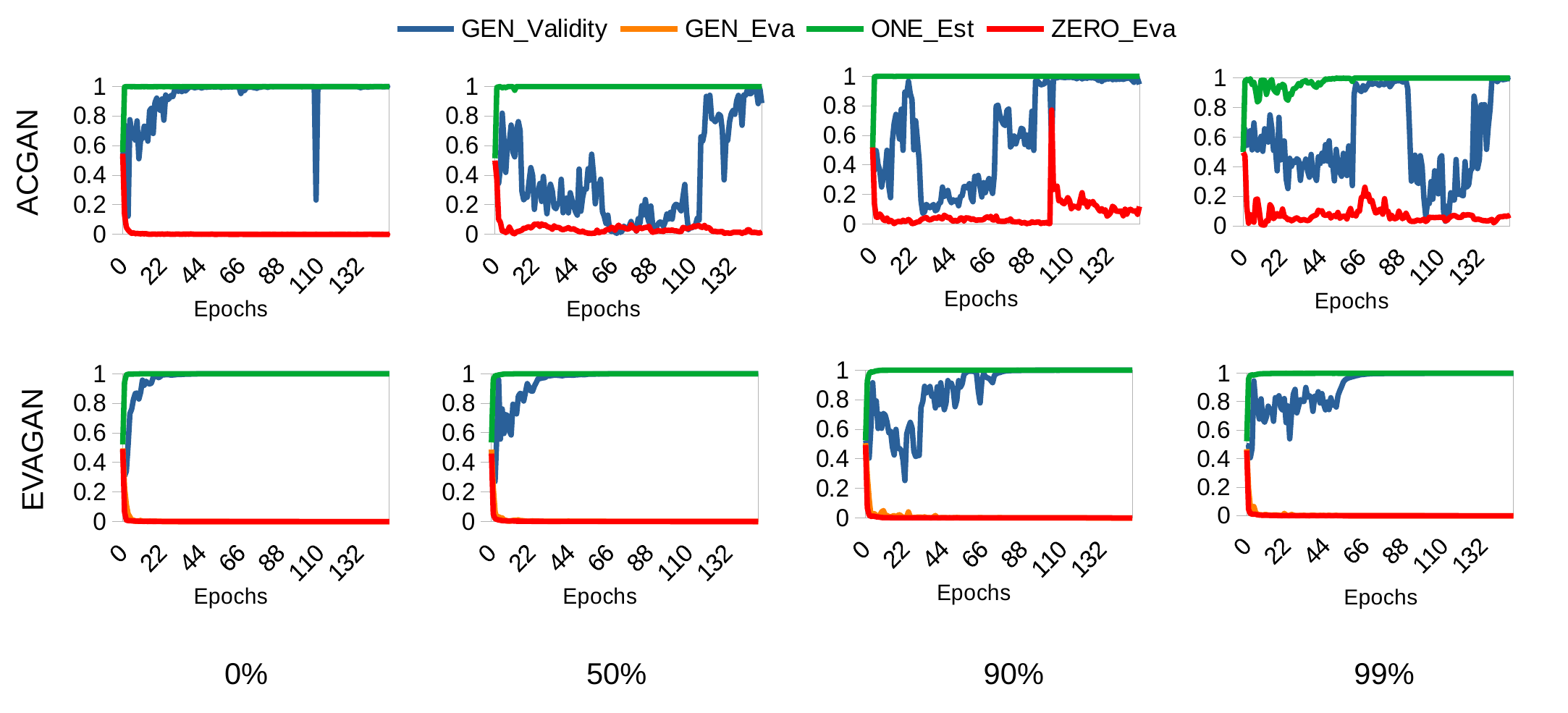}
\caption{CV GANs estimations: The estimations on the test set for ACGAN and EVAGN for MNIST dataset in different undersampling scenarios. The range of the estimation value on the y-axis is from 0 to 1}
\label{fig: estimations_cv}
\end{figure*}

For ACGAN\_CV and EVAGAN\_CV, we use MNIST handwritten digits dataset. Only two classes of digits, '0' and '1', were used in ACGAN\_CV, as due to SCCE, its model does not accept fewer than two classes. For ECAGAN\_CV, we use only the '0' digit as the minority class. Since the MNIST data is already balanced, we need to undersample the values of the '0' digit class to demonstrate the difference in performance. Four different undersampling levels have been devised in section \ref{sec: discussion}. The evaluation parameters were equivalent to those used in CC GANs. For instance, GEN\_Validity is the same as GEN\_VALIDITY; GEN\_Eva is similar to FAKE\_BOT\_EVA with the minority class from MNIST, i.e. '0' in our case. Similarly, ONE\_Est is equivalent to REAL\_NORMAL\_EST, and ZERO\_Eva is comparable to REAL\_BOT\_EVA in CC GANs. Figure \ref{fig: estimations_cv} demonstrates the quantitative results for the four undersampling scenarios. Note that out of four, the first scenario exhibits 0\% undersampling. There are three scenarios with undersampling, 50\%, 90\% and 99\%. For qualitative analysis, the output from $\mathcal{G}$s of both ACGAN\_CV and EVAGAN\_CV have been demonstrated in Figures \ref{fig: visual_1} and \ref{fig: visual_2}. These results are also based on the undersampling cases.

\begin{figure*}[tb!]
\centering
\includegraphics[width=12.8cm, height=6cm]{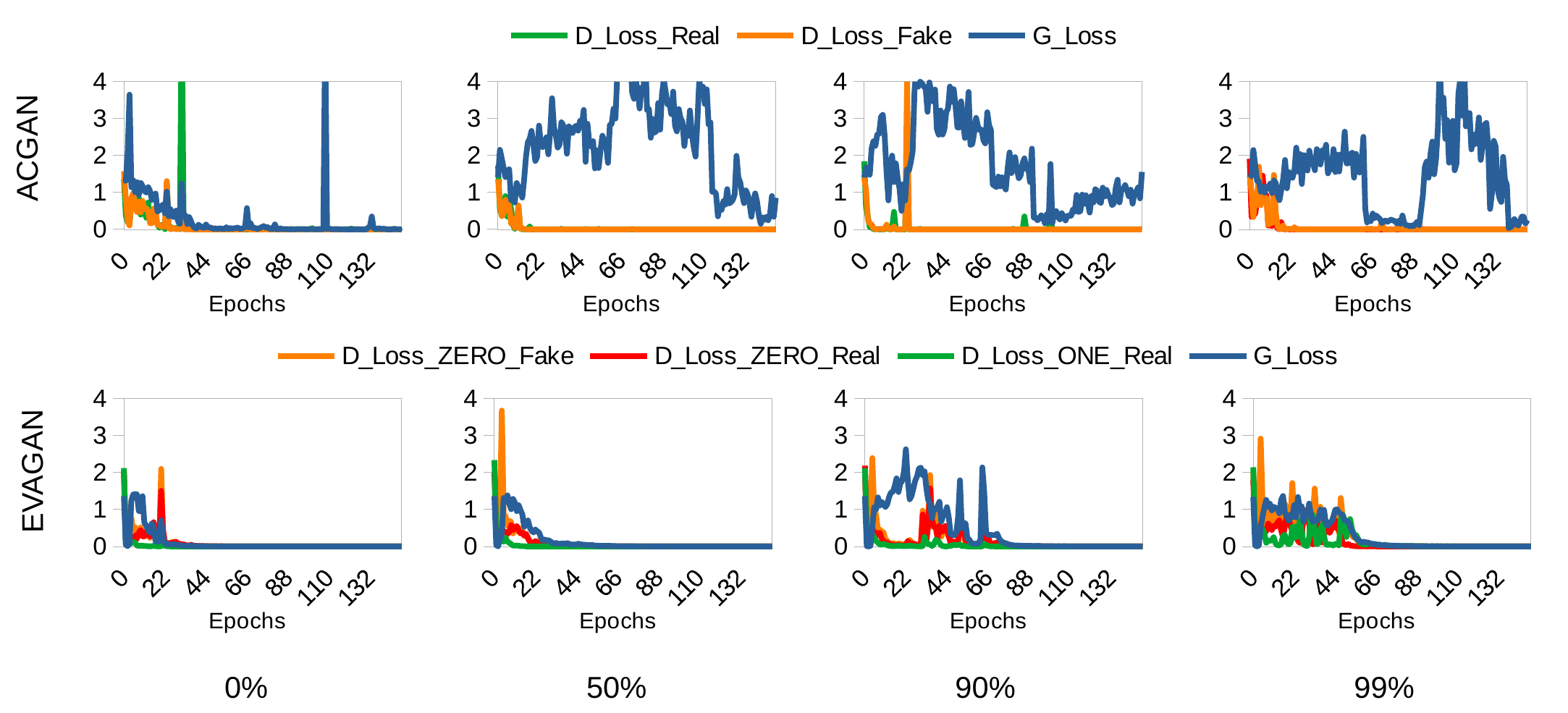}
\caption{CV GANs losses: The training losses on train set for ACGAN and EVAGN for MNIST dataset in different undersampling scenarios. The upper limit to the loss has been fixed to 4 for the sake of consistency to highlight the difference.}
\label{fig: losses_cv}
\end{figure*}

\section{Performance Comparison of EVAGAN with ACGAN}
\label{sec: discussion}

\subsection{Detection Performance}
In Figure \ref{fig: estimations_cc}, for the ACGAN\_CC, the values for the REAL\_NORMAL\_EST and REAL\_BOT\_EVA remain close to each other. This implies that the $\mathcal{D}$ of ACGAN\_CC is not able to discriminate between the majority and minority classes well due to the imbalance problem in all the three CC datasets. The $\mathcal{D}$ of ACGAN\_CC remains confused for the two classes in ISCX-2014 and CIC-2017 datasets. For the majority class, ACGAN\_CC performs equally well as EVAGAN\_CC for CIC-2018 (shown in the second row of Figure \ref{fig: estimations_cc}). However, due to the small number, it regards the minority class samples as the majority class instances. The second row in Figure \ref{fig: estimations_cc} shows the results of EVAGAN\_CC for the estimations on the test set. It can be observed that as compared to ACGAN\_CC, the $\mathcal{D}$ of EVAGAN\_CC perfectly differentiates between the majority and minority classes and, after each epoch, tends to improve its detection performance for all the three CC datasets. 

We have used FAKE\_BOT\_EVA as an indicator of evasion awareness of the $\mathcal{D}$ in the case of EVAGAN\_CC only because ACGAN\_CC generates two classes of data, so the $\mathcal{G}$ of ACGAN\_CC would generate a random number of samples from both classes leading to non-deterministic values of FAKE\_BOT\_EVA. However, we compare the performance of this metric with ML classifiers. The last row of Figure \ref{fig: estimations_cc} shows the results of the six different ML classifiers for the values of the majority, minority and generated class samples. It can be inferred that EVAGAN\_CC tends to outperform the ML classifiers for all three values after a certain number of epochs. The ML classifiers for black-box testing perform worst in the case of FAKE\_BOT\_EVA as compared to EVAGAN\_CC for all the three CC datasets. This implies that the $\mathcal{D}$ of EVAGAN\_CC is not only adept at discriminating between real minority samples but can also easily detect the fake minority samples that ML classifiers are not good at discerning. Another significant advantage of this $\mathcal{D}$ is that we do not need to employ ML classifiers in CC for learning adversarial evasion. Researchers use GANs to generate adversarial samples to be augmented with the training set for retraining ML classifiers to make them adversarially aware. In the case of EVAGAN\_CC, we save that time as the $\mathcal{D}$ classifier/detector model is trained alongside the GAN training.

It can be further illustrated from Figure \ref{fig: estimations_cc} that the value of GEN\_VALIDITY in the case of ACGAN\_CC seems to remain close to 0.5 for all the three CC datasets. It means that the $\mathcal{D}$ is confused in deciding whether the generated samples from $\mathcal{G}$ are real or fake. However, in the case of EVAGAN\_CC, for all the three datasets, $\mathcal{G}$'s performance is improving with each epoch. This implies that $\mathcal{D}$ is being fooled and still learning, while in the case of ACGAN\_CC, the $\mathcal{D}$ has already been saturated because $\mathcal{G}$ is not generating new samples that can fool $\mathcal{D}$. 

\subsubsection{CV GANs}
Figure \ref{fig: estimations_cv} demonstrated the results of different undersampling scenarios to mimic the low data regimes for the MNIST dataset. Note that the detection performance of the $\mathcal{D}$ for both ACGAN\_CV and EVAGAN\_CV for the majority and minority classes remains ideal from the very start. The reason is that, unlike CC datasets, the CV dataset has many strong features due to which $\mathcal{D}$ is easily able to differentiate between the '0' digit and '1' digit samples. However, the effect of undersampling can be seen for the minority class or digit '0' data. In contrast, the $\mathcal{D}$ of EVAGAN\_CV seems to be smart enough to give steady values for all the undersampling cases, especially for minority class evasion (as depicted in red colour lines). Due to the sufficient number of samples, the majority class should be detected easily by both GANs. However, in the case of 99\% undersampling, ACGAN\_CV exhibits a poor performance even detecting this class. For GEN\_Validity (represented by the blue lines), Figure \ref{fig: estimations_cv} shows that in undersampling cases, the performance of $\mathcal{G}$ of ACGAN\_CV deteriorates in the worst manner and does not show any useful pattern of learning. This implies that in low data regimes, $\mathcal{G}$ is not performing any better as compared to EVAGAN\_CV. However, EVAGAN\_CV also shows the deterioration in $\mathcal{G}$'s performance, but that is not as phenomenal as that of ACGAN\_CV.

\subsection{Stability}
The Figure \ref{fig: losses_cc} shows the $\mathcal{D}$ and $\mathcal{G}$ losses for CC GANs. It can be inferred from this diagram that the values for all the losses seem to be converging. This shows that the GANs are saturating towards Nash equilibrium. However, in the case of EVAGAN\_CC, the losses tend to be more steady with each epoch and achieve the lowest point sooner than ACGAN\_CC. 
Similarly, for CV GANs, the EVAGAN\_CV losses in all the undersampling cases tend to be more stable as compared to ACGAN\_CV as demonstrated in Figure \ref{fig: losses_cv}.

\begin{figure*}[tb!]
\centering
\includegraphics[width=18cm, height=8cm]{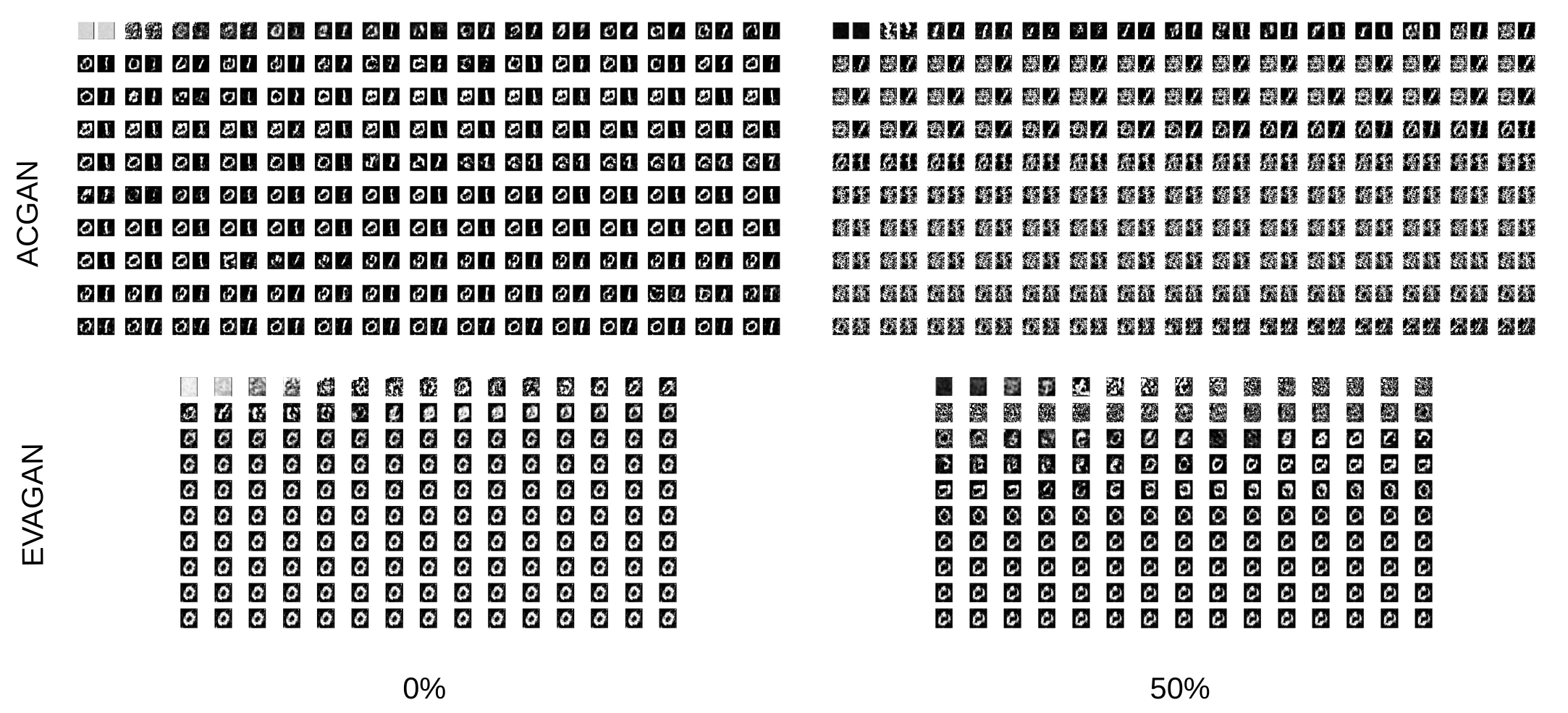}
\caption{Qualitative analysis of $\mathcal{G}$ for CV GANs with 0\% and 50\% undersampling of '0' digit class}
\label{fig: visual_1}
\end{figure*}

\begin{figure*}[tb!]
\centering
\includegraphics[width=18cm, height=8cm]{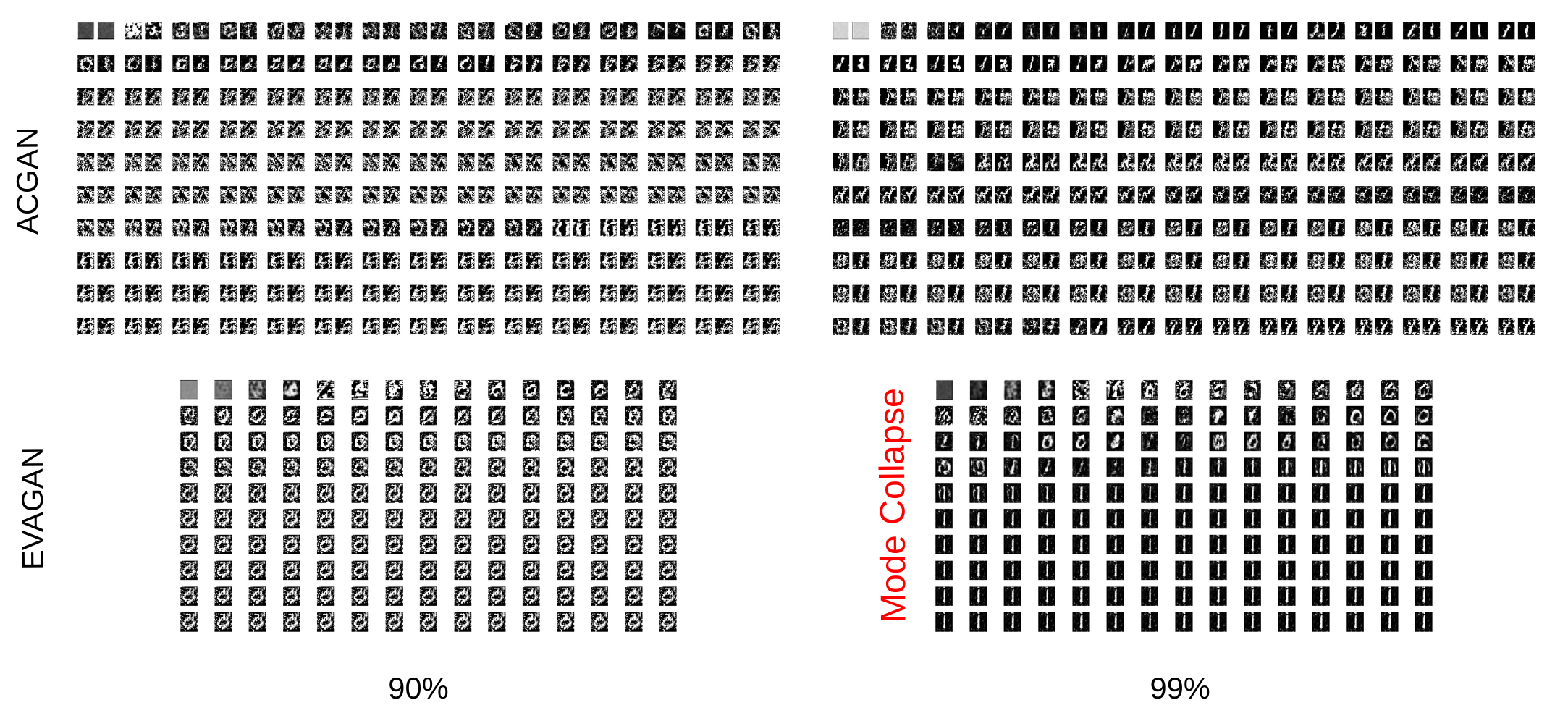}
\caption{Qualitative analysis of $\mathcal{G}$ for CV GANs with 90\% and 99\% undersampling of '0' digit class}
\label{fig: visual_2}
\end{figure*}

\subsection{Qualitative Performance}
It is non-trivial to demonstrate the performance of a GAN in the case of CC datasets \cite{guo2021ta}. Since we can not visualize the generated network traffic, we need to validate the EVAGAN with the help of CV datasets. The rationale for using CV datasets is that if EVAGAN outperforms ACGAN in unbalanced scenarios, it would be equally acceptable for CC datasets. Since our purpose is not to generate quality traffic for CC, we need to design an evasion-aware anomaly detector. So, evaluating EVAGAN\_CC for quality traffic generation is not within the scope of this work. 

The previously mentioned undersampling scenarios for CV GANs have been demonstrated in Figures \ref{fig: visual_1} and \ref{fig: visual_2}. There are two $15\times10$ matrices of pictures in each figure. The number of images in each matrix equals the total number of epochs, i.e. 150. In each figure, the upper row belongs to the ACGAN\_CV output of the $\mathcal{G}$ and the lower row corresponds to the output from $\mathcal{G}$ of EVAGAN\_CV. Note that for ACGAN\_CV, there are two classes being output from $\mathcal{G}$ and for EVAGAN\_CV, only one '0' digit class is generated. For the undersampling scenario, which contains 50\% fewer '0' class samples, the deterioration for ACGAN\_CV starts getting evident, but EVAGAN\_CV can generate '0' digits. For the case of 90\% undersampling, the ACGAN\_CV quality further deteriorates; however, EVAGAN\_CV is still generating the '0' class samples, although slightly faded. As expected, in the 99\% undersampling case, the ACGAN\_CV is still struggling to generate the minority class digit '0', but an interesting case has happened for EVAGAN\_CV. Since the number of samples is minuscule so the feedback taken from the $\mathcal{D}$ by $\mathcal{G}$, on some accidentally generated '1' digit,  gave a small value of $G\_Loss$. Due to this reason, the $\mathcal{G}$ started generating the majority class '1' digit after epoch 47. This situation is called a mode collapse, an inherent problem in GANs. However, we can infer that EVAGAN\_CV may not perform well in a highly unbalanced scenario. This is an interesting research direction to investigate further using other CV datasets. On the other hand, ACGAN\_CV is also stuck in mode collapse after epoch 140, where in place of class '0', the '1' class samples start appearing. However, in the case of EVAGAN\_CV, despite mode collapse, the generated samples from class '1' are of higher quality which means that its $\mathcal{G}$ is more powerful as compared to that of ACGAN in highly unbalanced scenarios.

\subsection{Time Complexity}
\begin{figure}[tb!]
\centering
\includegraphics[width=5.71cm, height=4cm]{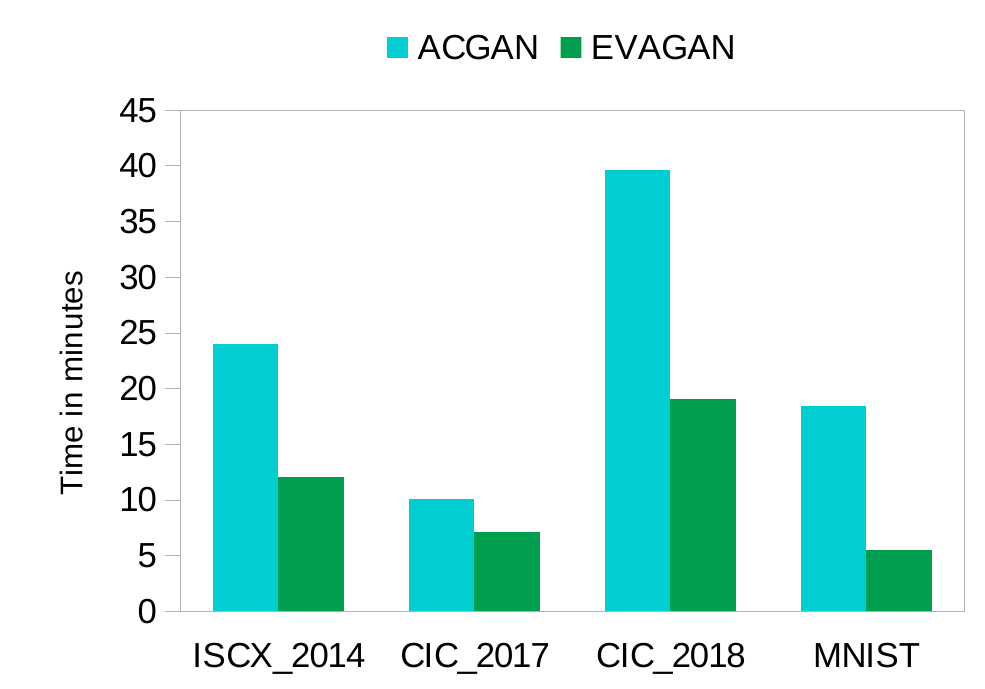}
\caption{Time complexity}
\label{fig: timeComplexity}
\end{figure}

The time complexity bar chart has been demonstrated in Figure \ref{fig: timeComplexity} where the y-axis represents the values of the training time in minutes. The MNIST dataset case with no undersampling was used to compare the results. The time complexity may vary on different platforms (for instance, Google Colab); however, the plot in Figure \ref{fig: timeComplexity} shows the results on the workstation that we have used (as mentioned in section \ref{sec: ImplementationDetails}). It can be observed that EVAGAN always takes less time than its counterpart for all four datasets. The reason lies in the notion that the $\mathcal{G}$ of EVAGAN in the cases of all the datasets needs to follow lesser diversity as compared to ACGAN. Although the batch size of 256 (given in Table \ref{table: GANs_hyper}) is the same for both GANs, the amount of time taken by EVAGAN is always less. Due to the stochastic nature of the input noise $z$ for the $\mathcal{G}$, we can not estimate the exact time in minutes for every training cycle; however, the average time of EVAGAN always remains less as compared to ACGAN. 

A question might arise why we did not make ACGAN generate only the minority class samples. The answer to this question is that we would have to make changes in the structure of both $\mathcal{G}$ and $\mathcal{D}$ along with the loss functions. The SCCE loss does not allow us to use less than two classes, so we need to use BCE loss with other structural modifications. EVAGAN is the name of this transformation.  

\section{Comparison of EVAGAN with Peer Techniques}
The EVAGAN model is an enhanced version of ACGAN, dedicated to low data regimes for learning adversarial evasion examples generated during GAN training. So the most suitable existing model for the comparison can be ACGAN, the details of which have been explained in section \ref{sec: discussion} previously. However, this section mentions peer techniques similar to EVAGAN that indirectly address the adversarial evasion problem. We have summarized the comparison in the following subsections and then in the form of a Table \ref{table: peerWorks}.

\subsection{Data Augmentation}
There are several techniques both in CV and CS that propose the data augmentation for enhancing the ML classifiers' performance \cite{usama2019generative, shahriar2020g, lee2019ae, abou2020evaluation, cheng2021packet, abdelaty2021gadot, qui2021strengthening, sabeel2021cvae } . However, EVAGAN itself acts as a powerful adversarial evasion-aware model in which the discriminator ($\mathcal{D}$) acts as a classifier. So there is no need to generate evasion samples from a GAN model, augment with the training set and then train a separate ML classifier. This property of EVAGAN makes it superior to all the techniques based on data augmentation in terms of time complexity.
\subsection{Computer Vision vs Cybersecurity Low Data Regimes}
There are plenty of works that address the problem of low data regimes \cite{li2022multi, frid2018gan, yi2019generative, waheed2020covidgan, chinbat2022ga3n, jo2022obgan, madhu2022envgan}. However, their datasets and model architectures differ from those used in this work. We have mentioned a few in Table \ref{table: peerWorks}. Since the ML studies are biased towards data, experimenting with other datasets can be a potential future work. 
\subsection{Architecture Comparison}
Authors in \cite{yin2019enhancing} proposed a model in which the discriminator is acting as a multiclass classifier; however, their work is not destined towards adversarial evasion generation in low data regimes as they are considering the normal class samples to train the generator of their GAN. Our work differs in a way that we do not feed our generator ($\mathcal{G}$) with normal class samples, which makes the $\mathcal{G}$'s job easier. This saves the training time and improves the estimation accuracy of malicious samples, even being scanty.

\subsection{Accuracy and Time Complexity}
EVAGAN produces ideal results of estimation for both majority and minority classes, as high as 100\% for all the datasets used, as mentioned in section \ref{sec: results}. The comparison has been provided with ACGAN; however, the accuracy in comparison with other similar models is at par as well. The accuracy values determined from the literature for some other models addressing similar problems have been given in Table \ref{table: peerWorks}. The time complexity as compared to the ACGAN model has been discussed in section \ref{sec: results} however; it would be non-trivial to compare with other peer models in respect of training time as the model architecture and hyperparameters vary enormously. The experiments for EVAGAN and ACGAN were performed on the same machine as mentioned in section \ref{sec: ImplementationDetails}, so we claim the time complexity comparison with ACGAN only.

\begin{table*}[ht!]

\centering
\caption{Comparison with Peer Works}

\label{table: peerWorks}

\centering
\begin{center}
\begin{tabular}{ | c | c | c |c |c |c |c |}
    
      \hline
    
      \thead{\textbf{Paper}} & \thead{\textbf{Addressing} \\ \textbf{Evasion} \\ \textbf{Problem}} & \thead{\textbf{Adversarial} \\ \textbf{Training/} \\\textbf{Augmentation}} & \thead{\textbf{Low} \\ \textbf{Data Regime}} & \thead{\textbf{Architecture}} & \thead{\textbf{Datasets Used}} & \thead{\textbf{Maximum} \\ \textbf{Accuracy}}  \\
      
      \hline
      
      ID-GAN \cite{yin2019enhancing} & \xmark& \xmark& \cmark& multiclass ACGAN& NSL-KDD &  83.10\%\\
      \hline
      
      G-IDS \cite{shahriar2020g} & \xmark& \cmark& \cmark& vanilla GAN& NSL-KDD&  -\\
      \hline
      
      AE-CGAN \cite{lee2019ae} & \xmark& \cmark& \cmark&  \makecell{Auto Encoder \\ with Conditional \\GAN}& CIC-2017&  100\%\\
      \hline
      
      \cite{abou2020evaluation} & \cmark& \cmark& \xmark& ANN, CNN, RNN& UNSW-NB15, NSL-KDD &  97\%\\
      \hline
      
      Attack-GAN \cite{cheng2021packet} & \cmark& \cmark& \xmark& Sequence GAN& CTU-13&  -\\
      \hline
      
      GADoT \cite{abdelaty2021gadot} & \cmark& \cmark& \xmark& WGAN-GP& \makecell{Custom-SYN,  Scapy-SYN, \\ CICIDS2017,  UNB201X} &  -\\
      \hline
      
      DIGFuPAS \cite{qui2021strengthening} & \cmark& \cmark& \cmark& WGAN& CICIDS2017&  -\\
      \hline
      
      min-max Training \cite{grierson2021min} & \cmark& \cmark& \cmark& DNN& NSL-KDD&  93.4\%\\
      \hline
      attackGAN \cite{zhao2021attackgan} & \cmark& \xmark& \cmark& WGAN& NSL-KDD&  -\\
      \hline
      
      CVAE-AN \cite{sabeel2021cvae} & \cmark& \cmark& \cmark& \makecell{Conditional VAE \\and GAN}& CICIDS2017 &  98\%\\
      \hline
      
      
      
     CEGAN \cite{suh2021cegan} & \xmark& \cmark& \cmark& CNN& \makecell{MNIST, EMNIST, \\ F-MNIST CIFAR-10,\\ CINIC-10}&  96.48\%\\
      \hline
      
      \textbf{EVAGAN} & \cmark& \xmark& \cmark&  binary class ACGAN & \makecell{ISCX-2014, CIC-2017, \\CIC-2018, MNIST}&  100\%\\
      \hline

   \end{tabular}

 \end{center}

 \end{table*}

\section{Conclusion}
\label{sec: conclusion}
Adversarial evasion attacks on AI-based systems are a portending threat that needs to be dealt with using intuitive methods. Adversarial learning is one of the modern techniques to make ML classifiers proactively adept at detecting adversarial evasion samples. This paper proposes a novel GAN model called EVAGAN that generates adversarial evasions in low data regimes. EVAGAN is an enhancement of a well-known model called ACGAN. EVAGAN aims to design an adversarial-aware classifier for anomaly detection. We have used two datasets; one from the cybersecurity domain for botnets and the other from the computer vision called MNIST. EVAGAN's discriminator is superior to ACGAN in terms of detection performance, stability, and time complexity. At the same time, the qualitative analysis shows that EVAGAN outperforms ACGAN in unbalanced scenarios. EVAGAN model has been designed for binary classification problems.

Further investigation for multiclass design is a potential research direction. Experiments with other datasets would be highly desirable to further evaluate EVAGAN for the said parameters. For the qualitative analysis, handwritten digits other than '0' and '1' could be used to validate EVAGAN's superiority over ACGAN. A comparison with few-shot learning could be an interesting research direction.



\bibliographystyle{IEEEtran}
\bibliography{ref}

\begin{IEEEbiography}[{\includegraphics[width=1in,height=1.25in,clip,keepaspectratio]{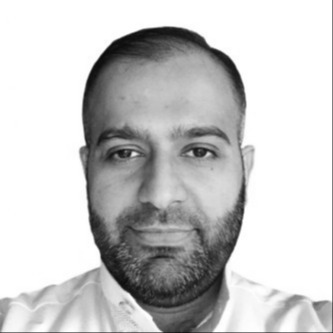}}]{Rizwan Hamid Randhawa} received a BS degree in Electronic Engineering from International Islamic University Islamabad, Pakistan in 2008 and Master in Computer Science from Information Technology University, Lahore, Pakistan in 2017. He has vast experience with embedded systems in Pakistan's private and public sector organisations. He is pursuing his PhD in Computer Science from Northumbria University, Newcastle upon Tyne, UK. His research interests include AI-based botnet detection, IoT Security and Embedded Systems Design \& Development for IoT platforms.

\end{IEEEbiography}

\begin{IEEEbiography}[{\includegraphics[width=1in,height=1.25in,clip,keepaspectratio]{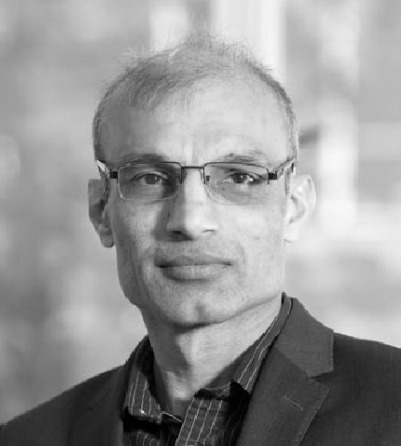}}]{Nauman Aslam} is a Professor in the Department of Computer and Information Science, Northumbria University, UK. Before joining Northumbria University as a Senior Lecturer in 2011, he worked as an Assistant Professor at Dalhousie University, Canada. He received his PhD in Engineering Mathematics from Dalhousie University, Canada, in 2008. Dr Nauman leads the Network Systems and Security research group at Northumbria University. His research interests cover diverse but interconnected areas related to communication networks. His current research focuses on addressing wireless body area networks and IoT, network security, QoS-aware communication in industrial wireless sensor networks, and Artificial Intelligence (AI) application in communication networks. He has published over 100 papers in peer-reviewed journals and conferences. Dr Nauman is a member of IEEE.
\end{IEEEbiography}

\begin{IEEEbiography}[{\includegraphics[width=1in,height=1.25in,clip,keepaspectratio]{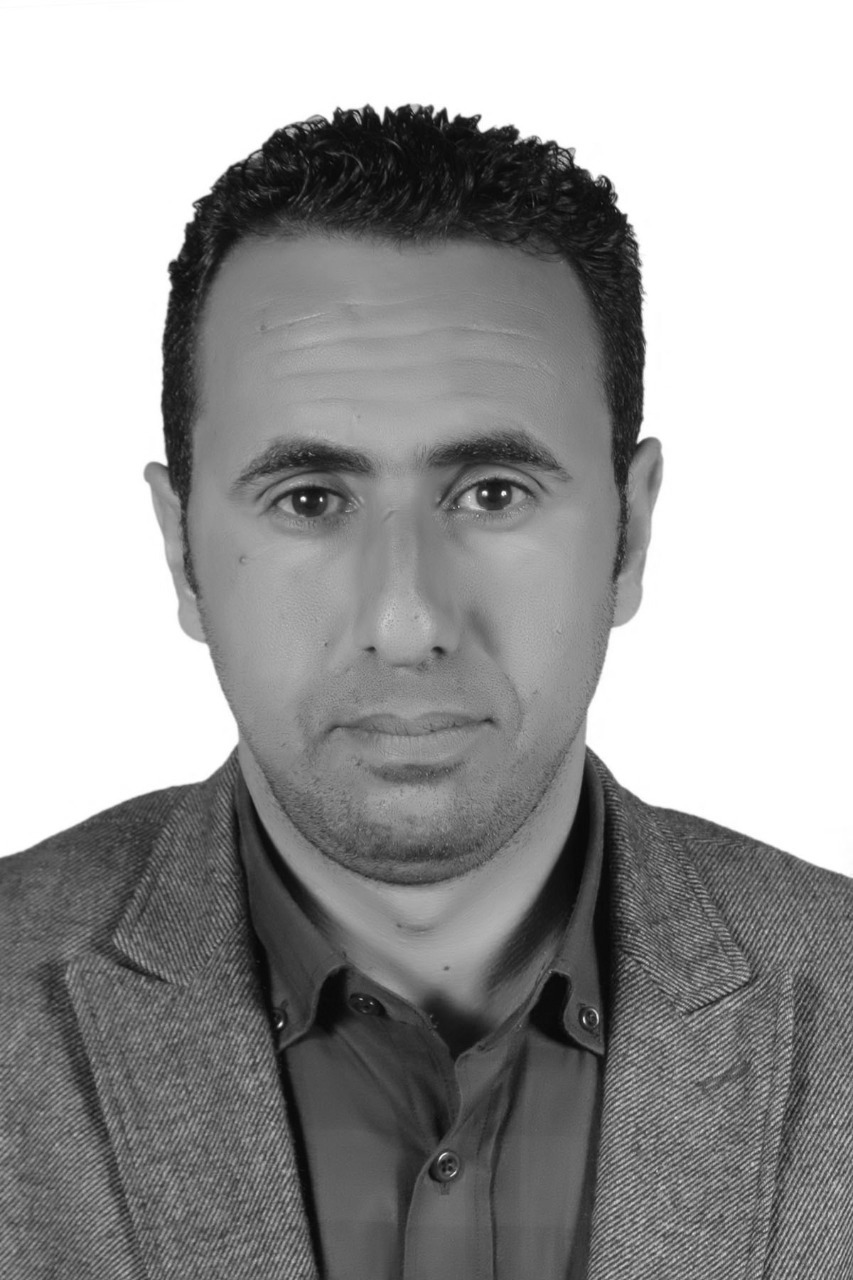}}]{Mohammed Alauthman} received his PhD degree in Network Security \& Botnet Detection from Northumbria University Newcastle, the UK, in 2016. He received a B.Sc. degree in Computer Science from Hashemite University, Jordan, in 2002, and an M.Sc. degree in Computer Science from Amman Arab University, Jordan, in 2004. Currently, he is Assistant Professor at the Information Security Department, Petra University, Jordan. His research interests include Cyber-Security, Cyber Forensics, Advanced Machine Learning and Data Science applications.
\end{IEEEbiography}

\begin{IEEEbiography}[{\includegraphics[width=1in,height=1.25in,clip,keepaspectratio]{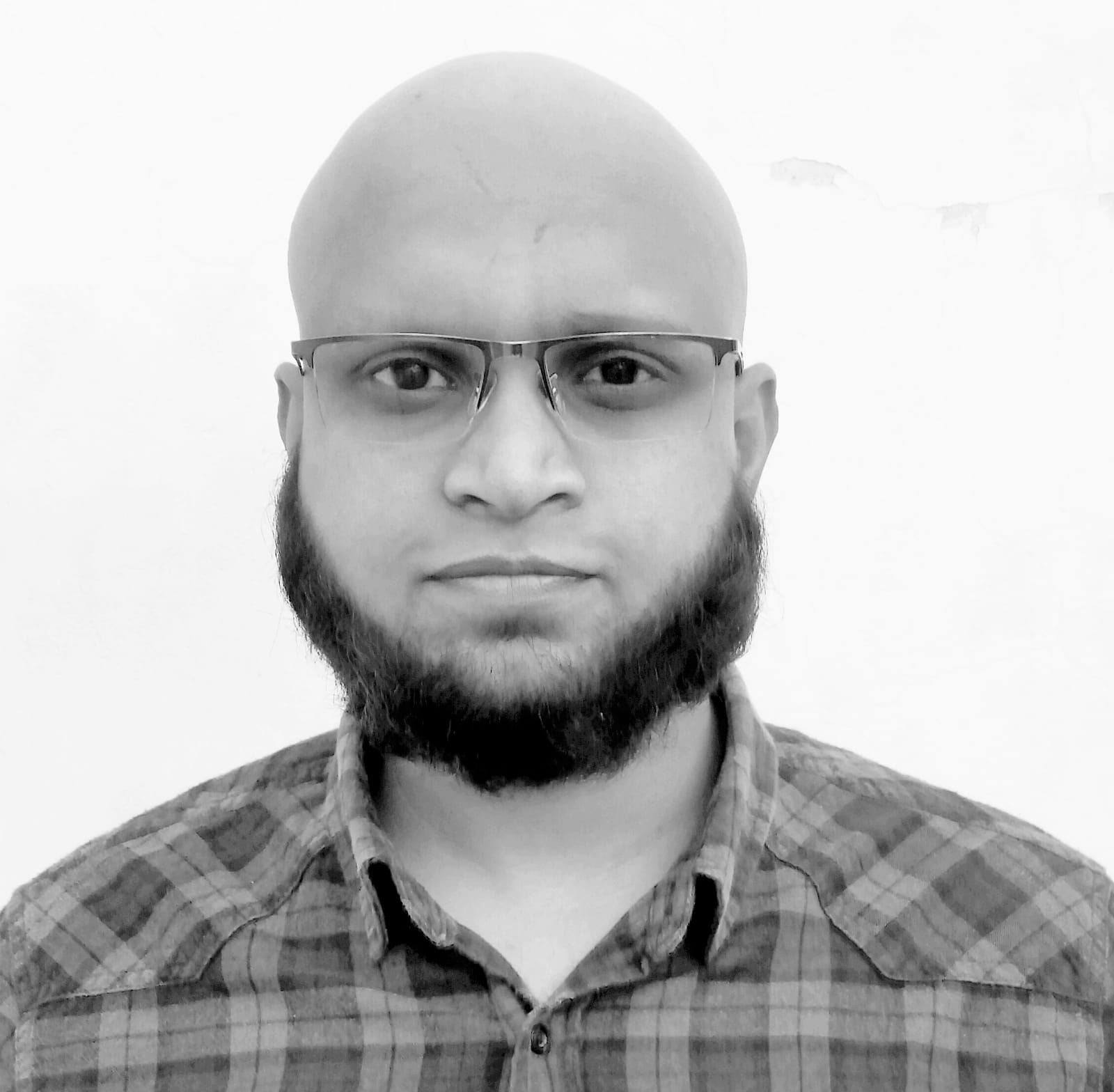}}]{Husnain Rafiq} received the B.S. and M.S. degrees in Computer Science from the Capital University of Science and Technology, Islamabad, Pakistan in 2015 and 2017, respectively. From 2015 to 2018, he was a Jr. Lecturer with the Capital University of Science and Technology, Islamabad, Pakistan. He is pursuing a PhD from Northumbria University Newcastle upon Tyne, UK in Android Malware Detection. His area of research includes Information Security and Forensics, Machine learning and Malware Analysis. 
\end{IEEEbiography}

\end{document}